\documentclass[aps,twocolumn,prd,preprintnumbers,superscriptaddress,nofootinbib,floatfix,balancelastpage,10pt]{revtex4-2}
\usepackage[bitstream-charter]{mathdesign}
\usepackage{XCharter}
\usepackage[dvipsnames,x11names]{xcolor}
\usepackage[colorlinks=true,
	        linkcolor=MathBlue,
	        citecolor=MathBlue,
	        filecolor=MathBlue,
	        urlcolor=MathBlue
	        ]{hyperref}
\usepackage[T1]{fontenc}
\usepackage[utf8]{inputenc}
\usepackage[british]{babel}
\usepackage{amsmath}
\usepackage{tikz}
\usepackage{orcidlink}
\usepackage{slashed}
\usepackage{placeins}
\usepackage[compat=1.1.0]{tikz-feynhand}
\usepackage{comment}
\usepackage{xfrac}
\usepackage[protrusion=true,expansion,kerning=true,tracking=true,final]{microtype}
\usepackage{tikz}
\usetikzlibrary{shapes.geometric}
\usepackage{lipsum}
\usepackage{braket}
\usepackage{tcolorbox}
\usepackage{soul}

\DeclareSymbolFont{usualmathcal}{OMS}{cmsy}{m}{n}
\DeclareSymbolFontAlphabet{\mathcal}{usualmathcal}

\definecolor{MathBlue}{rgb}{0.368417, 0.506779, 0.709798}
\definecolor{MathYellow}{rgb}{0.880722, 0.611041, 0.142051}
\definecolor{MathGreen}{rgb}{0.560181, 0.691569, 0.194885}
\definecolor{MathRed}{rgb}{0.922526, 0.385626, 0.209179}
\definecolor{MathViolet}{rgb}{0.528488, 0.470624, 0.701351}
\definecolor{LightGray}{gray}{0.91}
\definecolor{LightBlue}{rgb}{0.87, 0.94, 1}

\newcommand{\eq}[1]{\eqref{eq:#1}}
\newcommand{\fig}[1]{Fig.~\ref{fig:#1}}

\newcommand{\Sec}[1]{Sec.~\ref{sec:#1}}

\begin{document}

\title{Revisiting the \texorpdfstring{$\phi^6$}{phi**6} Theory in Three Dimensions at Large \texorpdfstring{$N$}{N}}%
\preprint{}

\author{Sandra~Kvedarait\.e\,\orcidlink{0000-0002-0269-9543}}
\email{skvedaraite@ugr.es}
\affiliation{Departamento de F\'isica Te\'orica y del Cosmos, Universidad de Granada, Campus de Fuentenueva, E–18071 Granada, Spain}

\author{Tom~Steudtner\,\orcidlink{0000-0003-1935-0417}}
\email{tom2.steudtner@tu-dortmund.de}
\affiliation{Fakult\"at Physik, Technische Universit\"at Dortmund, D-44221 Dortmund, Germany}

\author{Max~Uetrecht\,\orcidlink{0000-0001-8685-2543}}
\email{max.uetrecht@tu-dortmund.de}
\affiliation{Fakult\"at Physik, Technische Universit\"at Dortmund, D-44221 Dortmund, Germany}

\begin{abstract}
We investigate the $O(N)$--symmetric $\phi^6$ theory in three spacetime dimensions using dimensional regularisation and minimal subtraction. The predictions of other methods are scrutinised in a large-$N$ expansion.
We show how the tricritical line of fixed point emerges in a strict $N\to\infty$ limit but argue that it is not a physical manifestation.
For the first time in this explicit manner, we compute the effective potential at next-to-leading order in the $1/N$-expansion and discuss its stability.
The Bardeen-Moshe-Bander phenomenon is also analysed at next-to-leading order, and we demonstrate that it disappears without breaking the scale invariance spontaneously. Our findings indicate that the UV fixed point found by Pisarski persists at large $N$.
\end{abstract}

\maketitle
\tableofcontents

\section{Introduction}

In this work we reinvestigate a theory of $N$ real scalar fields exhibiting an $O(N)$ symmetry in $d=3$ spacetime dimensions. The most general renormalisable action is given by the Lagrangian
\begin{equation}\label{eq:lag}
	\begin{aligned}
		\mathcal{L} &=  \frac12 \partial_\mu\phi_k \partial^\mu\! \phi_k  - \frac{m^2}2 \phi^2
		- \frac\lambda{4!}\phi^4  - \frac\eta{6!} \phi^6\,,
	\end{aligned}
\end{equation}
where $\phi^2 \equiv \phi_k \phi_k$ with $k = 1,\dots,\,N$. In three dimensions, the coupling $\eta$ is classically marginal, while both $m^2$ and $\lambda$ have positive mass dimension.
\begin{figure}[ht]
	\centering
	\includegraphics[scale=.6]{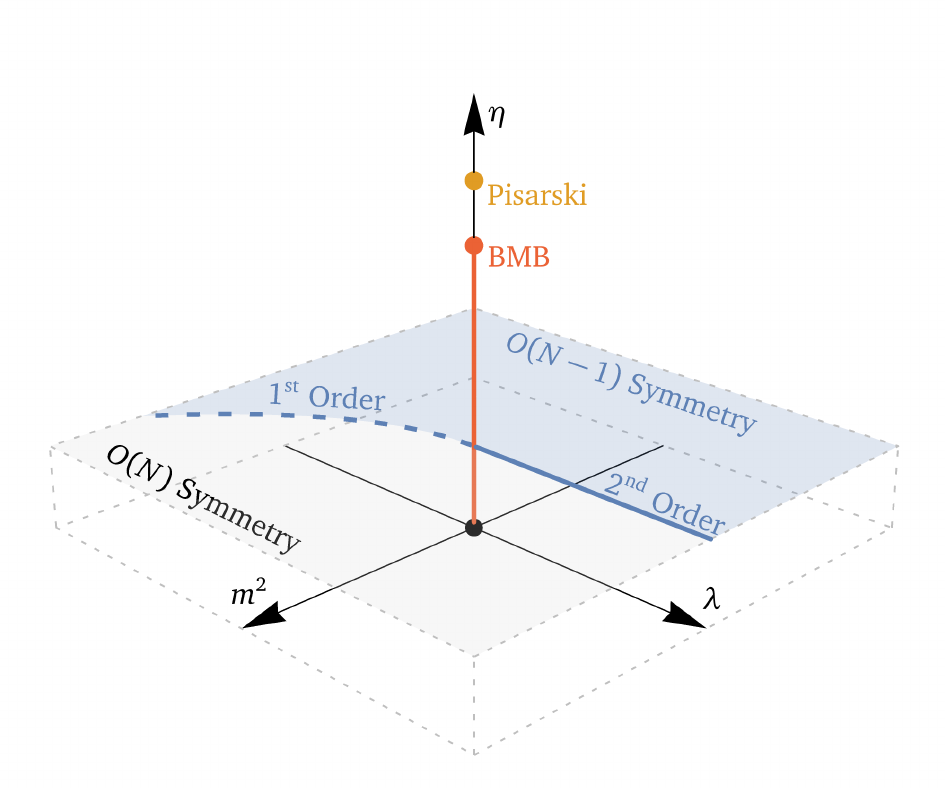}
	\caption{Schematic overview of the phase diagram for the theory~\eq{lag}. On a slice of a fixed value of $\eta$, the unbroken $O(N)$--symmetric phase (light grey surface) and broken $O(N-1)$ symmetry phase (light blue) are shown at tree level. 
	A second-order phase transitions occurs at $\lambda > 0$, $m^2=0$ (solid blue line) and a first-order one at $\lambda < 0$, $m^2 = \tfrac{5}{8} \lambda^2/\eta$ (dashed blue line). 
	Taking quantum corrections into account, a tricritical line of fixed points for $m^2 = \lambda = 0$ and $\eta > 0$ is reported~\cite{David:1984we,David:1985zz,Litim:2017cnl} in a strict $N \to \infty$ limit (red line).
	However, the line terminates at the BMB endpoint (red)~\cite{Bardeen:1983rv}.
	Beyond that line, there is a potential UV fixed point (yellow) found in a $1/N$-expansion~\cite{Townsend:1976sy,Appelquist:1981sf,Pisarski:1982vz,Hager:2002uq}. 
	}
	\label{fig:phases}
\end{figure}
The model~\eq{lag} plays a key role in understanding tricritical phenomena, such as those occurring in ${}^3$He$-$${}^4$He mixtures~\cite{PhysRevA.4.1071,AragaodeCarvalho:1976nt} or the nematic to smectic-$A$ transition in binary liquid crystal mixtures~\cite{ALBEN19731783,PhysRevLett.52.208}. 
Moreover, \eq{lag} is a rare example of  quantum field theories (QFTs) in which a non-trivial UV fixed point can be reliably identified using perturbative methods~\cite{Townsend:1976sy,Appelquist:1981sf,Pisarski:1982vz}. Other such instances include purely fermionic models in three~\cite{Gawedzki:1985ed,Rosenstein:1988pt,Gat:1990xi}, and certain gauge-Yukawa theories in four spacetime dimensions~\cite{Litim:2014uca,Litim:2015iea,Bond:2019npq,Litim:2023tym,Steudtner:2024pmd}.

In the following, we are interested in the limit where $N$ is very large, allowing for the extraction of non-perturbative results through a $1/N$-expansion.
An analysis of the phase structure has been conducted long ago, see e.g.~\cite{Townsend:1975kh,Townsend:1976sy,Townsend:1976sz,Appelquist:1981sf,Appelquist:1982vd}, 
and is displayed in \fig{phases} in the tree-level approximation.
The interaction potential is bounded from below at large field values if $\eta > 0$. 
Its $O(N)$ symmetry (light grey surface for fixed $\eta$) 
may spontaneously break down to $O(N-1)$ (light blue surface) 
via a second-order phase transition at $m^2 < 0$ (solid blue line) or a first order one for $m^2 > 0$ and $\lambda < - \sqrt{ 8\eta/5 }\,|m|  < 0$ (dashed blue line).\footnote{Note that transition orders are exchanged in the convention of~\cite{Townsend:1975kh}.} 
Along the line of the first-order phase transition, the $O(N)$-- and $O(N-1)$--symmetric phases coexist, culminating in a tricritical phase boundary at the intersection with the line of second-order transitions ($\eta$ axis of \fig{phases}).

In this work, we are mostly interested in this phase boundary at $m^2 = \lambda = 0$ for positive values of $\eta > 0$. 
There are two accounts of the large-$N$ UV behaviour of this theory that appear contradictory at a first glance.
On the one hand, Ref.~\cite{Townsend:1976sy,Appelquist:1981sf,Pisarski:1982vz} identified a single UV fixed point using perturbation theory.
In particular, Pisarski~\cite{Pisarski:1982vz} demonstrated that this fixed point is guaranteed to exist in a systematic $1/N$-expansion (see also~\cite{Hager:2002uq,Shrock:2023vkn}).
On the other hand, it has been argued that in a strict $N\to\infty$ limit, all points $m^2 = \lambda = 0$ exhibit scale invariance~\cite{David:1984we,David:1985zz,Litim:2017cnl}. This suggests a tricritical 
line of UV fixed points for all values $\eta \geq 0$ (red line in \fig{phases}).  
Ref.~\cite{Bardeen:1983rv} showed that this line terminates due to the Bardeen-Moshe-Bander (BMB) phenomenon, 
which generates a mass scale through non-perturbative effects.
This picture is corroborated using saddle point techniques and auxiliary fields in cut-off schemes, see 
e.g.~\cite{Bardeen:1983rv,David:1984we,Amit:1984ri,Gudmundsdottir:1984rr,Gudmundsdottir:1984vyf,David:1985zz,Omid:2016jve}
as well as the Functional Renormalisation Group~(FRG), see e.g.~\cite{Morris:1997xj,Litim:2017cnl,Yabunaka:2017uox,Litim:2018pxe,Fleming:2020qqx}.
Pisarski's fixed point (yellow in \fig{phases}) lies beyond the BMB endpoint where scale invariance is broken. 
Therefore, these non-perturbative arguments appear to suggest that the fixed point does not exist.
However, the line of fixed points along with the BMB endpoint disappears as $1/N$-corrections to the strict $N\to\infty$ limit are considered~\cite{David:1984we,David:1985zz,Amit:1984ri,Yabunaka:2017uox,Fleming:2020qqx}.
Pisarski's solution on the other hand survives $1/N$-corrections, but it is unclear if non-perturbative effects still eliminate the fixed point as they do in the $N\to\infty$ limit.
In addition, the methods and schemes used to access the tricritical line, BMB phenomenon and Pisarski's fixed point are quite different and often opaque on a technical level. 
This makes it difficult to draw a consistent picture about the large-$N$ limit.

In this work, we approach the problem from an accessible, purely diagrammatic angle.
Contrary to most previous works, we employ dimensional regularisation and the modified minimal subtraction scheme~\cite{Bollini:1972bi,Bollini:1972ui,tHooft:1973mfk,Bardeen:1978yd}. 
This allows to make better contact with the high-loop corrections of the renormalisation group equations~\cite{Pisarski:1982vz,Hager:2002uq}.
We evaluate the $\beta$-function and effective potential in a consistent $1/N$ expansion which involves resumming large families of diagrams in perturbation theory.

With these tools at hand, we investigate the disappearance of the tricritical line, push the expansion of the effective potential to next-to-leading order, and study the implications of the BMB phenomenon.

The remainder of this work is structured as follows:
\Sec{RGE} reviews Pisarski's fixed point and establishes a controlled large-$N$ expansion. In \Sec{Veff}, we argue that the tricritical line is spurious and only exists as an artefact of the $N \to \infty$ limit.
We also compute the effective potential at next-to-leading order. 
To the best of our knowledge, this is the first complete result of its kind in the literature.
The BMB phenomenon is scrutinised in \Sec{BMB}, and we demonstrate how it is replaced by stability as $1/N$ corrections are considered. Finally, we collect some concluding remarks in \Sec{Conclusion}.

%\cite{Yabunaka:2018mju,Yabunaka:2021fow,Yabunaka:2023jlf}?

\section{Renormalisation Group Equations}\label{sec:RGE}

We first consider the renormalisation group (RG) evolution in the large-$N$ limit. Note that for dimensional regularisation in $d=3-2\varepsilon$, 
odd loop orders do not exhibit UV divergencies~\cite{McKeon:1992cs}. 
Thus, $\beta$ and $\gamma$ functions in (modified) minimal subtraction schemes only receive contribution at even orders, starting at two-loop.  
The $\overline{\text{MS}}$ $\beta$-function of the marginal scalar sextic coupling $\eta$ has been computed by Pisarski at two and four loops~\cite{Pisarski:1982vz} and later extended to six loops in~\cite{Hager:2002uq}.
Retaining only the leading-$N$ terms at each loop order, it reads
\begin{equation}\label{eq:Pisarski-beta}
	\beta_\eta = \frac{N \eta^2}{5 (4\pi)^2} - \frac{\pi^2 N^3\eta^3}{7200(4\pi)^4}  +  c_6 \frac{N^4 \eta^4}{(4\pi)^6}  + \mathcal{O}(\eta^5)\,,
\end{equation}
where $c_6$ is a six-loop constant independent of $N$, see~\cite{Hager:2002uq}. We have verified \eq{Pisarski-beta} explicitly up to four-loop order.  
Analysing the coefficients in \eq{Pisarski-beta}, we find that in order to absorb the leading powers of $N$ into the coupling $\eta$ and obtain a finite $\beta$-function, 
the sextic coupling should be rescaled~\cite{Appelquist:1981sf} as 
\begin{equation}\label{eq:eta-hat}
	\hat{\eta} \equiv  \frac{N^{3/2} \eta}{  (4\pi)^2}\,, \qquad \text{ while } \qquad  \epsilon \equiv \frac{1}{\sqrt{N}} 
\end{equation}
can be treated as a small expansion parameter.
This means that only the four-loop term is leading, while both two- and six-loop contributions are suppressed by a factor of~$\epsilon$.
In fact, the only diagrams capable of generating the leading-$N$ contributions are~\cite{Pisarski:1982cm}\footnote{We are indebted to Ian Jack and Hugh Osborn to point out missing 8-loop diagrams. } 
\begin{equation}\label{eq:eta-large-N}
	\beta_\eta = \underbrace{\begin{tikzpicture}[baseline=.5em,scale=.8]
		\draw[fill=black] (-1.5em,0) circle (0.15em and 0.15em);
		\draw[fill=black] (+1.5em,0) circle (0.15em and 0.15em);
		\draw[fill=black] (0,2.414em) circle (0.15em and 0.15em);
		\draw[fill=none, line width=0.05em] (-1.5em,0) arc (120:60:+3.0em);
		\draw[fill=none, line width=0.05em] (-1.5em,0) arc (-120:-60:+3.0em);
		\draw[fill=none, line width=0.05em] (-1.5em,0) arc (180:120:+3.0em);
		\draw[fill=none, line width=0.05em] (-1.5em,0) arc (-60:0:+3.0em);
		\draw[fill=none, line width=0.05em] (1.5em,0) arc (-120:-180:+3.0em);
		\draw[fill=none, line width=0.05em] (1.5em,0) arc (0:60:+3.0em);
		\draw[fill=none, line width=0.05em] (-1.5em,0) -- (-2.5em,0.0em);
		\draw[fill=none, line width=0.05em] (-1.5em,0) -- (-1.75em,-0.75em);
		\draw[fill=none, line width=0.05em] (1.5em,0) -- (2.5em,0.0em);
		\draw[fill=none, line width=0.05em] (1.5em,0) -- (1.75em,-0.75em);
		\draw[fill=none, line width=0.05em] (0,2.414em) -- (0.75em,2.914em);
		\draw[fill=none, line width=0.05em] (0,2.414em) -- (-0.75em,2.914em);
	\end{tikzpicture}}_{\propto\  N^3 \eta^3} +
	\underbrace{
		\begin{tikzpicture}[baseline=-1.em,scale=.8]
		\draw[fill=black] (-1.5em,0) circle (0.15em and 0.15em);
		\draw[fill=black] (+1.5em,0) circle (0.15em and 0.15em);
		\draw[fill=black] (0,2.414em) circle (0.15em and 0.15em);
		\draw[fill=black] (-1.5em,-3em) circle (0.15em and 0.15em);
		\draw[fill=black] (+1.5em,-3em) circle (0.15em and 0.15em);
		\draw[fill=none, line width=0.05em] (-1.5em,0) arc (120:60:+3.0em);
		\draw[fill=none, line width=0.05em] (-1.5em,0) arc (-120:-60:+3.0em);
		\draw[fill=none, line width=0.05em] (-1.5em,0) arc (180:120:+3.0em);
		\draw[fill=none, line width=0.05em] (-1.5em,0) arc (-60:0:+3.0em);
		\draw[fill=none, line width=0.05em] (1.5em,0) arc (-120:-180:+3.0em);
		\draw[fill=none, line width=0.05em] (1.5em,0) arc (0:60:+3.0em);
		\draw[fill=none, line width=0.05em] (0,2.414em) -- (0.75em,2.914em);
		\draw[fill=none, line width=0.05em] (0,2.414em) -- (-0.75em,2.914em);
		\draw[fill=none, line width=0.05em] (-1.5em,-3em) arc (120:60:+3.0em);
		\draw[fill=none, line width=0.05em] (-1.5em,-3em) arc (-120:-60:+3.0em);
		\draw[fill=none, line width=0.05em] (-1.5em,-3.0em) -- (-2.5em,-3.0em);
		\draw[fill=none, line width=0.05em] (-1.5em,-3.0em) -- (-1.75em,-3.75em);
		\draw[fill=none, line width=0.05em] (1.5em,-3.0em) -- (2.5em,-3.0em);
		\draw[fill=none, line width=0.05em] (1.5em,-3.0em) -- (1.75em,-3.75em);
		\draw[fill=none, line width=0.05em] (-1.5em,0) arc (-210:-150:+3em);
		\draw[fill=none, line width=0.05em] (-1.5em,0) arc (-330:-390:+3em);
		\draw[fill=none, line width=0.05em] (+1.5em,0) arc (-210:-150:+3em);
		\draw[fill=none, line width=0.05em] (+1.5em,0) arc (-330:-390:+3em);
	\end{tikzpicture}
	+ 
	\begin{tikzpicture}[baseline=-1.em,scale=.8]
		\draw[fill=black] (-1.5em,0) circle (0.15em and 0.15em);
		\draw[fill=black] (+1.5em,0) circle (0.15em and 0.15em);
		\draw[fill=black] (0,2.414em) circle (0.15em and 0.15em);
		\draw[fill=black] (-1.5em,-3em) circle (0.15em and 0.15em);
		\draw[fill=black] (+1.5em,-3em) circle (0.15em and 0.15em);
		\draw[fill=none, line width=0.05em] (-1.5em,0) arc (180:120:+3.0em);
		\draw[fill=none, line width=0.05em] (-1.5em,0) arc (-60:0:+3.0em);
		\draw[fill=none, line width=0.05em] (1.5em,0) arc (-120:-180:+3.0em);
		\draw[fill=none, line width=0.05em] (1.5em,0) arc (0:60:+3.0em);
		\draw[fill=none, line width=0.05em] (0,2.414em) -- (0.75em,2.914em);
		\draw[fill=none, line width=0.05em] (0,2.414em) -- (-0.75em,2.914em);
		\draw[fill=none, line width=0.05em] (-1.5em,-3em) arc (120:60:+3.0em);
		\draw[fill=none, line width=0.05em] (-1.5em,-3em) arc (-120:-60:+3.0em);
		\draw[fill=none, line width=0.05em] (-1.5em,-3em) to [bend left=10] (+1.5em,-3em);
		\draw[fill=none, line width=0.05em] (-1.5em,-3em) to [bend right=10] (+1.5em,-3em);
		\draw[fill=none, line width=0.05em] (-1.5em,-0em) -- (-2.5em,-.0em);
		\draw[fill=none, line width=0.05em] (-1.5em,-0em) -- (-1.8em,+1.0em);
		\draw[fill=none, line width=0.05em] (1.5em,-0em) -- (2.5em,-.0em);
		\draw[fill=none, line width=0.05em] (1.5em,-0em) -- (1.8em,+1.0em);
		\draw[fill=none, line width=0.05em] (-1.5em,0) arc (-210:-150:+3em);
		\draw[fill=none, line width=0.05em] (-1.5em,0) arc (-330:-390:+3em);
		\draw[fill=none, line width=0.05em] (+1.5em,0) arc (-210:-150:+3em);
		\draw[fill=none, line width=0.05em] (+1.5em,0) arc (-330:-390:+3em);
	\end{tikzpicture}
	+ 
	\begin{tikzpicture}[baseline=-1.em,scale=.8]
		\draw[fill=black] (-1.5em,0) circle (0.15em and 0.15em);
		\draw[fill=black] (+1.5em,0) circle (0.15em and 0.15em);
		\draw[fill=black] (0,2.414em) circle (0.15em and 0.15em);
		\draw[fill=black] (-1.5em,-3em) circle (0.15em and 0.15em);
		\draw[fill=black] (+1.5em,-3em) circle (0.15em and 0.15em);
		\draw[fill=none, line width=0.05em] (-1.5em,0) arc (180:120:+3.0em);
		\draw[fill=none, line width=0.05em] (-1.5em,0) arc (-60:0:+3.0em);
		\draw[fill=none, line width=0.05em] (1.5em,0) arc (-120:-180:+3.0em);
		\draw[fill=none, line width=0.05em] (1.5em,0) arc (0:60:+3.0em);
		\draw[fill=none, line width=0.05em] (0,2.414em) -- (0.75em,2.914em);
		\draw[fill=none, line width=0.05em] (0,2.414em) -- (-0.75em,2.914em);
		\draw[fill=none, line width=0.05em] (-1.5em,-3em) to [bend left=20] (1.5em,0em);
		\draw[fill=none, line width=0.05em] (-1.5em,-3em) to [bend right=20] (1.5em,0em);
		\draw[fill=none, line width=0.05em] (+1.5em,-3em) to [bend left=20] (-1.5em,0em);
		\draw[fill=none, line width=0.05em] (+1.5em,-3em) to [bend right=20] (-1.5em,0em);
		\draw[fill=none, line width=0.05em] (-1.5em,-3.0em) -- (-2.5em,-3.0em);
		\draw[fill=none, line width=0.05em] (-1.5em,-3.0em) -- (-1.75em,-3.75em);
		\draw[fill=none, line width=0.05em] (1.5em,-3.0em) -- (2.5em,-3.0em);
		\draw[fill=none, line width=0.05em] (1.5em,-3.0em) -- (1.75em,-3.75em);
		\draw[fill=none, line width=0.05em] (-1.5em,0) arc (-210:-150:+3em);
		\draw[fill=none, line width=0.05em] (-1.5em,0) arc (-330:-390:+3em);
		\draw[fill=none, line width=0.05em] (+1.5em,0) arc (-210:-150:+3em);
		\draw[fill=none, line width=0.05em] (+1.5em,0) arc (-330:-390:+3em);
	\end{tikzpicture}
	}_{\propto\ N^6 \eta^5} +  \dots 
\end{equation}
and appear every four loop orders, while all other orders are suppressed. 
Keeping only leading-$N$ terms at each loop order, the rescaled $\beta$-function reads
\begin{equation}\label{eq:Pisarski-beta-hat}
	\beta_{\hat{\eta}} = \frac{ \epsilon}{5} \hat{\eta}^2 - \frac{\pi^2}{7200}  \hat{\eta}^3 + \epsilon  c_6 \hat{\eta}^4  + c_8 \hat{\eta}^5 + \dots \,.
\end{equation}
Due to the signs of two- and four-loop terms in \eq{Pisarski-beta-hat}, a non-trivial UV fixed point emerges~\cite{Pisarski:1982vz}
\begin{equation}\label{eq:Pisarski}
	\hat{\eta}^* = \frac{1440}{\pi^2} \epsilon + \mathcal{O}(\epsilon^3)\,.
\end{equation}
In the limit $N\to\infty$ ($\epsilon \to 0$), the fixed point becomes increasingly small and asymptotically free. The coupling $\hat{\eta}(\epsilon)$ can be systematically expanded as a power series in $\epsilon$, where each coefficient is completely determined by $\beta_{\hat{\eta}}$ up to a certain loop order. Thus, higher loops are increasingly suppressed and do not spoil the existence of the fixed point.
The coefficient $\propto \epsilon^2$ in~\eq{Pisarski} vanishes while next coefficient $\propto \epsilon^3$ requires both the six- and eight-loop parts of the $\beta$-function, the latter being currently unavailable. This is because the corresponding terms in~\eq{Pisarski-beta-hat} both give contributions $\mathcal{O}(\epsilon^5)$ at the fixed point. 

Furthermore, we provide the eigenvalues of the stability matrix $\vartheta_{1,2,3}$ and field anomalous dimension\footnote{
	We follow the convention $\phi_\text{bare} = \sqrt{Z_\phi} \phi$ and $\gamma_\phi = \tfrac12 \mathrm{d}\log Z_\phi / \mathrm{d}\log \mu$. 
	} $\gamma_\phi^*$ at the UV fixed point
\begin{equation}
	\begin{aligned}
		\vartheta_{1} &= - \frac{288}{\pi^2} \,\epsilon^2 + \mathcal{O}(\epsilon^4)\,, \\
		\vartheta_{2,3} &= \mp \frac{8 \sqrt{3}}{\pi} \, \epsilon + \mathcal{O}(\epsilon^3)\,, \\
		\gamma_\phi^* &= - \frac{96}{\pi^4} \, \epsilon^4 + \mathcal{O}(\epsilon^5)\,,
	\end{aligned}
\end{equation}
where $\vartheta_1$ corresponds to the $\hat{\eta}$ direction and is UV attractive, while $\vartheta_{2,3}$ are a mixture of the superrenormalisable couplings $m^2$ as well as $\hat{\lambda} \propto N \lambda$ revealing one attractive and one repulsive direction.

Since the coefficient $\propto \epsilon$ in \eq{Pisarski} is positive, we have $\hat{\eta}^* > 0$ which implies that the classical potential is bounded from below. In order to check if this stability persists at the quantum level, we calculate the effective potential next.

\section{Effective Potential}\label{sec:Veff}

In the following section, we compute the effective potential, which incorporates quantum corrections to the classical potential. 
We start out by briefly reviewing the formalism behind the effective potential~\cite{Coleman:1973jx,Jackiw:1974cv,Sher:1988mj} before proceeding with its determination in the large-$N$ expansion.
Technical aspects of this loop calculation are collected in the App.~\ref{sec:appendix}.

\subsection{Definition}

The effective potential $V_\text{eff}$ is defined as the generating functional of all one-particle irreducible (1PI), 
zero-momentum Green's functions.
Following the procedures outlined in~\cite{Coleman:1973jx,Jackiw:1974cv,Sher:1988mj}, we obtain the generating functional of all 1PI Green's functions 
$\Gamma$ by shifting the $N$-component scalar field $\phi = \braket{\phi} + \hat{\phi}$ around a constant background field $\braket{\phi}$, along with integrating out the quantum fluctuations, i.e.
\begin{equation}\label{eq:1pi_func_shifted}
	e^{i \Gamma[\braket{\phi}]}=\int \mathscr{D} \hat{\phi} e^{i \int \mathrm{d}^3 x[\mathcal{L}(\braket{\phi}+\hat{\phi})+J \hat{\phi}]} \,.
\end{equation}
By construction, taking the $n$th functional derivative of $\Gamma[\braket{\phi}]$ with respect to $\braket{\phi}$ and setting $\braket{\phi} = 0$ 
yields the corresponding $n$-point 1PI Green's function. All one-point tadpoles in $\hat{\phi}$ are subtracted by the source term $J = -\delta \Gamma[\braket{\phi}]/\delta \braket{\phi}$.

Performing the path integral in~\eqref{eq:1pi_func_shifted} leads to the 1PI effective action
\begin{equation}
	\Gamma[\braket{\phi}]=\int \mathrm{d}^3 x\left[-V_{\text {eff }}(\braket{\phi})+\mathcal{O}(\partial \braket{\phi})\right] \,,
\end{equation}
whose zero-momentum term is given by the effective potential $V_{\text {eff }}$. On a practical level, $\braket{\phi}$ can be assumed as constant and absorbed into masses and couplings of the quantum field $\hat{\phi}$. 
The effective potential is then obtained via
\begin{equation}\label{eq:V_eff_loop_exp}
	\begin{aligned}
		V_{\mathrm{eff}}(\braket{\phi})&=V_{\mathrm{cl}}(\braket{\phi}) -\frac{i}{2} \int \frac{\mathrm{~d}^3 k}{(2 \pi)^{3}} \log \left[\operatorname{det} i \mathscr{P}(k, \braket{\phi})\right] \\ 
		&\phantom{=\ }+\left\langle 0 \left|T \exp \left(i \int d^3 x \mathcal{L}_{\text {int }}\right)\right| 0\right\rangle \,.
	\end{aligned}
\end{equation}
Here, $V_{\mathrm{cl}}$ is the classical potential appearing in the Lagrangian, $\mathscr{P}$ 
is the inverse propagator of the quantum field $\hat{\phi}$, and $\mathcal{L}_{\text {int }}$ is the interaction Lagrangian containing cubic and higher self interactions in $\hat{\phi}$.
Diagrammatically, the tree-level term corresponds to the classical potential, the logarithmic determinant to the one-loop vacuum tadpole, and the last term comprises all two- and 
higher loop contributions.
Eq.~\eq{V_eff_loop_exp} allows for a systematic computation of $V_{\mathrm{eff}}$, equivalent to the standard perturbative loop expansion.
In general, the action may contain several scalars, each of which has to be decomposed into a background field and one or more modes of quantum fluctuations. 
Eq.~\eq{V_eff_loop_exp} generalises to
\begin{equation}\label{eq:Veff-1L-template}
	V_\text{eff} = V_\text{cl} - \frac{1}{12 \pi}\sum_{\hat{\phi}} m_{\hat{\phi}}^3 + V_\text{1PI}
\end{equation}
in three spacetime dimensions, where the second term sums over all scalar quantum fluctuations and the third term stems from all 1PI vacuum diagrams at two loops and higher. They correspond to the second and third terms in \eq{V_eff_loop_exp}.

\subsection{Large-N Limit}

Now, we compute the effective potential for the theory~\eqref{eq:lag} with vanishing mass and quartic coupling, i.e. $m = \lambda = 0$, at leading order (LO) in large-$N$.
We shift the the scalar field $\phi = \braket{\phi} + \hat{\phi}$ around a constant background field $\braket{\phi}$, 
which can be rotated into a single component $\varphi$ using the global symmetry transformations. The explicit construction
\begin{equation}\label{eq:phi-GH}
	\phi(x) = \big(\varphi + H(x),\, G_1(x),\dots,\,G_{N-1}(x)\,\big)^\intercal
\end{equation}
breaks the global symmetry down to $O(N-1)$, and the shifted quantum field $\hat{\phi}$ consists of a Higgs mode $H$ as well as a $(N-1)$-component 
Nambu--Goldstone mode $G$ with components labeled by $G_i$ with $i = 1,\dots,\,N-1$.
The single field $H$ does not contribute to the effective potential at LO in the large-$N$ limit. Only interactions of the $G_i$ are relevant, which are described by the potential
\begin{equation}\label{eq:V_G}
	V_G = \frac{\eta}{6!} \varphi^6 + \frac{1}{2} \frac{\eta \varphi^4}{120} G^2 + \frac{1}{4!} \frac{\eta \varphi^2}{10} G^4 + \frac{\eta}{6!} G^6\,,
\end{equation} 
with $G^2 \equiv G_i G_i$.
The first term represents the classical potential, the second is a mass term for the quantum field $G_i$, followed by a quartic and sextic interaction.
This leads to the LO effective potential 
\begin{equation}\label{eq:Veff-G}
	\begin{aligned}
		V_\text{eff}^\text{LO} &= \frac{\eta}{6!} \varphi^6 \left[ 1 - \frac{N}{4\pi} \sqrt{\frac{\eta}{30}}  \right] + V_\text{1PI}^\text{LO}\,,
	\end{aligned}
\end{equation}
where the second term arises from the one-loop contribution to~\eq{Veff-1L-template}, 
which sums over the masses of all real scalar modes in the theory, and $V_\text{1PI}^\text{LO}$ denotes all higher loop terms.
They are given by the tadpole diagrams 
\begin{equation}\label{eq:V1PI-LO}
	V_\text{1PI}^\text{LO} =  \tikz [baseline=0em] {
		\draw[fill=none, line width=0.05em] (8.3em,1.0em) circle (0.6em and 0.6em);
		\draw[fill=none, line width=0.05em] (8.3em,-.2em) circle (0.6em and 0.6em);
		\draw[fill=black] (8.3em,.4em) circle (0.15em and 0.15em);
	} + \tikz [baseline=0em] {
		\draw[fill=none, line width=0.05em] (7.2em,.40em) circle (0.6em and 0.6em);
		\draw[fill=none, line width=0.05em] (8.3em,1.0em) circle (0.6em and 0.6em);
		\draw[fill=none, line width=0.05em] (8.3em,-.2em) circle (0.6em and 0.6em);
		\draw[fill=black] (7.92em,.4em) circle (0.15em and 0.15em);
	} + \tikz [baseline=0.6em] {
		\draw[fill=none, line width=0.05em] (8.3em,2.2em) circle (0.6em and 0.6em);
		\draw[fill=none, line width=0.05em] (8.3em,1.0em) circle (0.6em and 0.6em);
		\draw[fill=none, line width=0.05em] (8.3em,-.2em) circle (0.6em and 0.6em);
		\draw[fill=black] (8.3em,.4em) circle (0.15em and 0.15em);
		\draw[fill=black] (8.3em,1.6em) circle (0.15em and 0.15em);
	} + \tikz [baseline=0.6em] {
		\draw[fill=none, line width=0.05em] (7.2em,.40em) circle (0.6em and 0.6em);
		\draw[fill=none, line width=0.05em] (8.3em,1.0em) circle (0.6em and 0.6em);
		\draw[fill=none, line width=0.05em] (8.3em,-.2em) circle (0.6em and 0.6em);
		\draw[fill=none, line width=0.05em] (8.3em,-.2em) circle (0.6em and 0.6em);
		\draw[fill=none, line width=0.05em] (8.3em,2.2em) circle (0.6em and 0.6em);
		\draw[fill=black] (7.92em,.4em) circle (0.15em and 0.15em);
		\draw[fill=black] (8.3em,1.6em) circle (0.15em and 0.15em);
	} + \tikz [baseline=0.6em] {
		\draw[fill=none, line width=0.05em] (8.3em,2.2em) circle (0.6em and 0.6em);
		\draw[fill=none, line width=0.05em] (8.3em,1.0em) circle (0.6em and 0.6em);
		\draw[fill=none, line width=0.05em] (8.3em,-.2em) circle (0.6em and 0.6em);
		\draw[fill=none, line width=0.05em] (9.5em,1.0em) circle (0.6em and 0.6em);
		\draw[fill=black] (8.3em,.4em) circle (0.15em and 0.15em);
		\draw[fill=black] (8.3em,1.6em) circle (0.15em and 0.15em);
		\draw[fill=black] (8.9em,1.0em) circle (0.15em and 0.15em);
	}  + \tikz [baseline=0.6em] {
		\draw[fill=none, line width=0.05em] (8.3em,2.2em) circle (0.6em and 0.6em);
		\draw[fill=none, line width=0.05em] (8.3em,1.0em) circle (0.6em and 0.6em);
		\draw[fill=none, line width=0.05em] (8.3em,-.2em) circle (0.6em and 0.6em);
		\draw[fill=none, line width=0.05em] (9.5em,-.2em) circle (0.6em and 0.6em);
		\draw[fill=black] (8.3em,.4em) circle (0.15em and 0.15em);
		\draw[fill=black] (8.3em,1.6em) circle (0.15em and 0.15em);
		\draw[fill=black] (8.9em,-.2em) circle (0.15em and 0.15em);
	}+  \dots \,.
\end{equation}

For the effective potential to remain finite in the large-$N$ limit, we proceed to discuss the large-$N$ scaling of the LO contributions~\eq{V1PI-LO}.
Consider an $\ell$-loop graph $\mathcal{G}$ consisting of $p$ propagators and including $n_6$ sextic as well as $n_4$ quartic interactions. Due to being a vacuum diagram,
the relation
\begin{equation}
	p = 3\, n_6 + 2\, n_4
\end{equation}
holds. Moreover, the mass dimensions of the potential implies
\begin{equation}
	[V] = 3 = 3\,\ell - 2\,p + n_4\,.
\end{equation}
Each of the  LO diagrams~\eq{V1PI-LO} fulfills
\begin{equation}
	\ell = 1 + n_4 + 2\,n_6.
\end{equation}
Overall, the leading-$N$ contribution of the graph reads
\begin{equation}\label{eq:Gscaling}
	\begin{aligned}
		\mathcal{G} & \propto N^\ell \eta^{n_6 + n_4 - p + 3\ell/2} \varphi^6 = N^\ell \eta^{1+\ell/2} \varphi^6 \,.
	\end{aligned}
\end{equation}
Note that~\eq{Gscaling} also holds  for the one-loop term in~\eq{Veff-G}.
Thus, the leading large-$N$ scaling of the effective potential remains the same for all loop orders, if the sextic coupling is rescaled via
\begin{equation}\label{eq:eta-bar}
	\bar{\eta} \equiv \frac{N^2 \eta}{(4\pi)^2 5!} \,,
\end{equation}
with additional factors for later convenience. This means that the effective potential is suppressed by an overall factor of $N^{-2}$. 

Before computing the LO effective potential~\eq{Veff-G}, let us address the apparent discrepancy between the leading-$N$ contributions to the $\beta$-function and the effective potential, 
which are captured by the rescalings \eq{eta-hat} and \eq{eta-bar}, respectively. 
To be precise, $\bar{\eta} = 1/120 \,\sqrt{N} \hat{\eta}$. 
It is no contradiction that both quantities require a different scaling of the coupling $\eta$ to be finite but non-zero in large-$N$ limit. 

The effective potential is obtained from the finite terms of 1PI vacuum graphs. At LO in large-$N$, all contributions stem from the tadpole graphs displayed in~\eq{V1PI-LO}, which warrant the rescaling~\eq{eta-bar}. 
On the other hand, the $\beta$-function is extracted from $1/\varepsilon$-poles in dimensional regularisation. The tadpoles~\eq{V1PI-LO} neither exhibit such poles, nor contribute as subgraphs to diagrams that do.
In mass-independent regularisation schemes, this can be readily understood: scalars can be treated as massless when computing UV poles, and tadpoles vanish altogether in this case.
Even in regularisation schemes where massless tadpoles do not vanish, they do not enter in the renormalisation of the sextic interaction, 
but rather of the scalar mass and quartic operator. This is evident in cutoff-schemes, where $\bar{\eta}$ does not require a counterterm, see e.g.~\cite{Townsend:1975kh}.
In consequence, leading-$N$ contributions to the $\beta$-function take the shape of~\eq{eta-large-N} which are captured by the rescaling~\eq{eta-hat}.

The RG evolution of $\bar{\eta}$ at $1/N$ is:
\begin{equation}\label{eq:eta-bar-beta}
	\beta_{\bar{\eta}} = \frac{24\bar{\eta}^2 - 2\pi^2 \bar{\eta}^3}{N} + \mathcal{O}(N^{-2}).
\end{equation} 
Thus, the UV fixed point still prevails
\begin{equation}\label{eq:Pisarski-bar}
	\bar{\eta}^* = \frac{12}{\pi^2} + \mathcal{O}(N^{-1}) \approx 1.2159 + \mathcal{O}(N^{-1})\,.
\end{equation}
As before, the leading correction of \eq{Pisarski-bar} is already determined with two- and four-loop $\beta$-functions, its $N^{-1}$ correction requires both six- and eight-loop RGEs.
Note that \eq{Pisarski-bar} does not appear to be under strict perturbative control as it approaches a finite value for $N\to\infty$. This is merely an artefact of the rescaling~\eq{eta-bar}, its existence is still guaranteed to all orders of the $\beta$-function, as demonstrated in~\Sec{RGE}.

In contrast, taking the na\"ive limit $N\to\infty$ in \eq{eta-bar-beta}  leads to the  conclusion that
the theory can be treated as conformal for all values of $\bar{\eta}$, revealing a line of fixed points that extends until it is 
broken by other means~\cite{Bardeen:1983st,Litim:2016hlb,Litim:2017cnl,Litim:2018pxe}.
This interpretation does not actually capture the leading large-$N$ dynamics~\eq{eta-large-N} of the RG evolution.
The line of fixed points is an artefact of employing a large-$N$ rescaling which renders the entire $\beta$-function subleading, and then dropping it entirely before analysing for fixed points.
In consequence, the line cannot be justified in a consistent $1/N$ expansion \eq{eta-bar-beta}, 
and breaks down as soon as subleading contributions are included to the $\beta$-functions~\cite{David:1984we,David:1985zz,Amit:1984ri}.
Furthermore, a continuous line in $\bar{\eta}$  cannot simply emerge at ``$N=\infty$'' as this would mean that the theory is non-interacting due to $\eta \propto \bar{\eta}/N^2 = 0$.

In~\cite{Yabunaka:2017uox,Fleming:2020qqx}, it has been argued that the line of fixed points can be approached in dimensional continuation around $d=3- \varepsilon$ via a combined $N\to\infty$ and $\varepsilon \to 0$ limit. In this case, two fixed-points 
\begin{equation}
	\bar{\eta}_{\mp}(\alpha) = \frac{6}{\pi^2} \left(1 \mp \sqrt{1 - \frac{\pi^2}{36} \alpha}\right) + \mathcal{O}(N^{-1})
\end{equation}
emerge~\cite{Pisarski:1982vz,Osborn:2017ucf} where 
\begin{equation}
	\alpha = \lim_{\substack{
	\ N\to\infty \\ 
	\varepsilon \to 0}} \varepsilon N \quad \text{ with } \quad 0 \leq \alpha \leq \frac{36}{\pi^2}
\end{equation}
is a free parameter which describes how the double limit is taken. Each valid choice of $\alpha$ corresponds to a fixed point, forming a line that lies asymptotically below three spacetime dimensions. However, the hard limit $d=3$ with large $N$ stipulates $\alpha=0$, where only the trivial $(\bar{\eta}_{-}(0) = 0)$ and UV fixed point $(\bar{\eta}_{+}(0) = \bar{\eta}^*)$ are retained.

Instead of a line of conformal fixed points, there is merely an accidental scale invariance in the LO large-$N$ contributions to the effective potential. Once subleading terms are included, this scale invariance is broken and only recovered at the UV fixed point~\eq{Pisarski-bar}.
An analogous picture should also emerge using the FRG when the expansion of the flow equation and the large-$N$ limit are interchanged. Note however that consistent results beyond leading-$N$ might require a departure from the local potential approximation.

Next, we devise a strategy to resum all LO loop diagrams~\eq{V1PI-LO} contributing to the effective potential. 
A direct summation of vacuum graphs~\eq{V1PI-LO} is complicated to generalise without double counting.
However, all LO contributions take the form of repeated tadpole insertions
\begin{tikzpicture}[baseline=-.3em]
	\draw[fill=none, line width=0.05em] (-1.0em,0) -- (0em,0);
	\draw[fill=none, line width=0.05em] (1.0em,0) -- (0em,0);
	\draw[fill=black] (0,0) circle (0.075em and 0.075em);
	\draw[fill=none] (0,0.3em) circle (0.3em and 0.3em);
\end{tikzpicture} and \begin{tikzpicture}[baseline=-.3em]
	\draw[fill=none, line width=0.05em] (-1.0em,0) -- (0em,0);
	\draw[fill=none, line width=0.05em] (1.0em,0) -- (0em,0);
	\draw[fill=black] (0,0) circle (0.075em and 0.075em);
	\draw[fill=none] (0,0.3em) circle (0.3em and 0.3em);
	\draw[fill=none] (0,-0.3em) circle (0.3em and 0.3em);
\end{tikzpicture} into scalar propagators. Thus, all LO terms can be resummed by introducing a dressed propagator \begin{tikzpicture}[baseline=-.3em]
	\draw[fill=none, line width=0.05em] (-1.0em,0) -- (0em,0);
	\draw[fill=none, line width=0.05em] (1.0em,0) -- (0em,0);
	\draw[fill=none, line width=0.05em] (-.15em,-0.15em) -- (+.15em,0.15em);
	\draw[fill=none, line width=0.05em] (+.15em,-0.15em) -- (-.15em,0.15em);
\end{tikzpicture} recursively defined via 
\begin{equation}\label{eq:gap-LO}
\begin{tikzpicture}[baseline=-.3em]
		\draw[fill=none, line width=0.05em] (-1.0em,0) -- (0em,0);
		\draw[fill=none, line width=0.05em] (1.0em,0) -- (0em,0);
		\draw[fill=none, line width=0.05em] (-.15em,-0.15em) -- (+.15em,0.15em);
		\draw[fill=none, line width=0.05em] (+.15em,-0.15em) -- (-.15em,0.15em);
	\end{tikzpicture} = \ \begin{tikzpicture}[baseline=-.3em]
		\draw[fill=none, line width=0.05em] (-1.0em,0) -- (0em,0);
		\draw[fill=none, line width=0.05em] (1.0em,0) -- (0em,0);
	\end{tikzpicture} + \begin{tikzpicture}[baseline=-.3em]
		\draw[fill=none, line width=0.05em] (-1.5em,0) -- (0,0);
		\draw[fill=none, line width=0.05em] (+1.5em,0) -- (0,0);
		\draw[fill=none, line width=0.05em] (0em,+.6em) circle (0.6em and 0.6em);
		\draw[fill=none, line width=0.05em] (-.15em,1.35em) -- (+.15em,1.05em);
		\draw[fill=none, line width=0.05em] (+.15em,1.35em) -- (-.15em,1.05em);
		\draw[fill=black] (0,0) circle (0.15em and 0.15em);
		\draw[fill=none, line width=0.05em] (+.8em,-0.15em) -- (+1.1em,+0.15em);
		\draw[fill=none, line width=0.05em] (+.8em,+0.15em) -- (+1.1em,-0.15em);
	\end{tikzpicture}  + \begin{tikzpicture}[baseline=-.3em]
		\draw[fill=none, line width=0.05em] (-1.5em,0) -- (0,0);
		\draw[fill=none, line width=0.05em] (+1.5em,0) -- (0,0);
		\draw[fill=none, line width=0.05em] (0em,-.6em) circle (0.6em and 0.6em);
		\draw[fill=none, line width=0.05em] (0em,+.6em) circle (0.6em and 0.6em);
		\draw[fill=none, line width=0.05em] (-.15em,1.35em) -- (+.15em,1.05em);
		\draw[fill=none, line width=0.05em] (+.15em,1.35em) -- (-.15em,1.05em);
		\draw[fill=none, line width=0.05em] (-.15em,-1.35em) -- (+.15em,-1.05em);
		\draw[fill=none, line width=0.05em] (+.15em,-1.35em) -- (-.15em,-1.05em);
		\draw[fill=black] (0,0) circle (0.15em and 0.15em);
		\draw[fill=none, line width=0.05em] (+.8em,-0.15em) -- (+1.1em,+0.15em);
		\draw[fill=none, line width=0.05em] (+.8em,+0.15em) -- (+1.1em,-0.15em);
	\end{tikzpicture}\quad .
\end{equation}
The tadpoles of \eq{gap-LO} factorise and are finite when integrating over loop momenta. 
Thus, their resummation is equivalent to introducing a dynamical mass parameter $M^2$ of the quantum field in addition 
to its mass $m^2 = 1/120 \, \eta \varphi^4$, recursively defined via a gap equation
\begin{equation}\label{eq:gap-mass-LO}
	M^2 = m^2 - \frac{N \eta \varphi^2 M}{240 \pi}  + \frac{N^2 \eta M^2}{1920 \pi^2}\,.
\end{equation}
We find two solutions~\cite{Townsend:1976sz,Matsubara:1984rk,Matsubara:1987iz}
\begin{equation}\label{eq:dynamical-mass}
	M_\pm = \frac{4\pi}{N} \frac{\varphi^{2}}{1 \pm \bar{\eta}^{-1/2}}\,.
\end{equation}
It is straightforward to verify that only $M_+$ corresponds to a perturbative expansion of the gap equation by 
recursively inserting \eq{gap-mass-LO} into its \textit{l.h.s.} up to a finite power of $m$, and comparing with \eq{dynamical-mass} 
expanded as a power series in $\bar{\eta}$. 
On the other hand, $M_-$ does not reproduce the perturbative expansion, and is negative for small values of $\bar{\eta}$, 
though the sign changes for $\bar{\eta} > 1$. 
Whether this solution is non-perturbative or rather a spurious shall be addressed later.

All leading large-$N$ quantum corrections are resummed by inserting the dynamical mass parameter 
$M_+$~\eq{dynamical-mass} in the propagator of the quantum field $G_i$ in place of the tree-level mass. Thus, the na\"ive 
expectation would be that all LO contributions to the effective potential~\eq{V1PI-LO} can be 
resummed by inserting $M_+$ into the one-loop term~\eq{Veff-1L-template} of the effective potential
\begin{equation}\label{eq:naive-LO}
	V^\text{LO}_\text{eff} \overset{?}{=} \frac{\eta}{6!} \varphi^6 - \frac{N}{12\pi} M_+^3
\end{equation}
and not include any explicit higher-loop diagrams. 
However, this approach does not reproduce the perturbative expansion of~\eq{V1PI-LO}. 
The 1PI effective potential is inadequate for resummation techniques that rely on insertion of bilinear operators, 
such as a dynamical mass term. The ansatz~\eq{Veff-1L-template} integrates out all tree-level bilinears and assumes 
that all two-loop and higher contributions are due to operators that are at least cubic in the field.
To compute the effective action, the diagrams in~\eq{V1PI-LO} either need to be resummed without relying on the dynamical mass, 
or the formalism needs to be extended to keep track of both $\braket{\phi}$ and bilinears $\braket{\phi^2}$ 
separately~\cite{Cornwall:1974vz}. We opt for the latter option, which we will review in the next section.

\subsection{Composite Operator Effective Action}
In this section, we review the formalism developed in~\cite{Cornwall:1974vz}. 
For ease of notation, we will work in an example theory with a single scalar field $\Phi$, 
for which we formulate an effective action $\Gamma[\phi,\,\chi]$ in terms of the classical field 
$\phi(x) \equiv \braket{\Phi(x)}$, but also the composite bilinear operator $\chi(x,y) \equiv \braket{T \Phi(x)\Phi(y)}$.
To this end, the generating functional of all connected Greens functions is defined with two source terms 
\begin{equation}
	\begin{aligned}
		e^{i W[J,\,K]} &= \int\mathcal{D}\Phi \exp\bigg\{ i \int \mathrm{d}^3 x \bigg[\mathcal{L}(\phi(x))   \\
		&+ J(x) \Phi(x) + \frac{1}{2} \int \mathrm{d}^3 y \,\Phi(x) K(x,y) \Phi(y) \bigg]\bigg\}	\,,
	\end{aligned}
\end{equation}
which is then shifted by a double Legendre transformation 
\begin{equation}
	\begin{aligned}
		&\Gamma[\phi,\chi]  = W[J,\,K] - \int\mathrm{d}^3x \, J(x) \phi(x)\\
		&\ - \frac{1}{2}\iint\mathrm{d}^3x\mathrm{d}^3y \left(\phi(x)\phi(y)+\chi(x,y)\right)K(x,y)\,,
	\end{aligned}
\end{equation}
yielding the effective action of all connected, \textit{two-particle irreducible} (2PI) Greens functions $\Gamma[\phi,\chi]$.
Roughly speaking, the effective action $\Gamma[\phi,\chi]$ is obtained by shifting scalar fields around a classical background $\Phi\mapsto\Phi + \phi$ as well as all connected two-point correlators by $\chi$ before integrating out $\Phi$.
Writing the original action as $S[\phi] = \int\mathrm{d}^3x \mathcal{L}(\phi(x))$, the effective action is obtained via 
	\begin{equation}\label{eq:Gamma-composite}
		\Gamma[\phi,\,\chi] = S[\phi] + \frac{i}{2} \mathrm{Tr}\left[\log \chi^{-1} + \frac{\delta^2 S}{\delta \phi^2} \chi \right]  + S_\text{2PI-int}\,,
\end{equation}
where $\mathrm{Tr}[\dots]$ includes a spacetime integration, but in general also a trace of global indices and different species of scalar fields.
$S_\text{2PI-int}$ contains all two-particle irreducible graphs at two-loop or higher with $\chi$ being used as a dressed propagator, and all cubic and higher order interaction vertices obtained after 
shifting $\Phi$ around $\phi$.
The one-particle irreducible effective action
\begin{equation}
	\Gamma[\phi] = \Gamma[\phi,\,\chi_0]
\end{equation}
is recovered by inserting $\chi = \chi_0$ at a stationary point of the action
\begin{equation}\label{eq:Gamma2PI-stationary}
	\frac{\delta\Gamma[\phi,\chi]}{\delta \chi} \bigg|_{\chi=\chi_0} = 0\,,
\end{equation}
which corresponds to classical solutions of the path integral such as ground states. Note that the condition \eq{Gamma2PI-stationary} is equivalent to a gap equation for the field $\Phi$.

Analogously to the effective potential of the scalar field $\phi$~\eq{V_eff_loop_exp}, the composite effective potential $V_{\mathrm{2PI}}$ 
is obtained as the generating functional of all \textit{zero-momentum} 2PI Green's functions, i.e.
\begin{equation}
	\Gamma[\phi,\,\chi] = \int  \mathrm{d}^3x  \left[-V_\text{2PI}(\phi, \chi) + \mathcal{O}\left(\partial \phi, \partial \chi\right) \right] \,.
\end{equation}

\subsection{Leading Order Potential}\label{subsec:LOV2PI}
We now apply the composite-operator formalism \eq{Gamma-composite} to compute the effective potential at leading order at large-$N$. 
As before, we employ the decomposition \eq{phi-GH}, which shifts $\phi$ around a constant background field $\varphi$. 
We also utilise that only the Nambu-Goldstone mode $G_i$ gives corrections at leading order in large-$N$, and that these can be resummed into a mass parameter $M$ for this field via \eq{gap-LO}. 
Thus, the formalism~\cite{Cornwall:1974vz} can be condensed into a more practical shape: instead of dealing with a dressed propagator, 
which also resums loop corrections that are explicitly momentum dependent, it is sufficient to keep track of a dynamical mass parameter $M$. 
Concretely,
\begin{equation}\label{eq:chi-G-LO}
	\chi^G_{ij}(x,y) = \braket{T G_i(x) G_j(y)} = \delta_{ij} \int \frac{\mathrm{d}^{d}p}{(2\pi)^{d}} \frac{e^{-ip(x-y)}}{p^2 - M^2}
\end{equation}
is a suitable ansatz for the dressed propagator. 

Using the vertex rules stemming from~\eq{V_G},
we compute \eq{Gamma-composite} and collect all non-derivative terms of the 2PI effective action in an effective potential $V_\text{2PI}$. 
The one-loop corrections, dimensionally regularised in $d=3-2\varepsilon$, read
\begin{equation}\label{eq:V2PI-1L}
	\begin{aligned}
		&\qquad \quad S[\phi] \Big|_{\varphi} \!= -\frac{\eta}{6!}\varphi^6 \int \!\!\!\mathrm{d}^3 x ,\\
		&\mathrm{Tr}\left[\log \chi^{-1}  \right] \Big|_{\varphi}  \!=  (N-1)\!\!\int \!\!\!\frac{\mathrm{d}^d p}{(2\pi)^d} \log\left(p^2 - M^2\right) \int \!\!\!\mathrm{d}^3 x, \\
		& \mathrm{Tr}\left[\frac{\delta^2 S}{\delta \phi^2} \chi \right] \Big|_{\varphi} = (N-1)\!\!\int\!\!\! \frac{\mathrm{d}^d p}{(2\pi)^d} \frac{p^2 - \tfrac1{120}\eta \varphi^4}{p^2 - M^2} \int\!\!\! \mathrm{d}^3 x,
	\end{aligned}
\end{equation}
when evaluated at $\phi(x) = \varphi$.
Thus, we obtain
\begin{equation}
		V_\text{2PI} = \frac{\eta}{6!} \varphi^6 - \frac{N M^3}{12\pi} - \frac{N M}{8\pi}\left( \frac{\eta \varphi^4}{120}  - M^2\right) + V_\text{2PI-int}\,, 
\end{equation}
where the first three terms correspond to the ones of \eq{Gamma-composite} in the same order. 
The last term $V_\text{2PI-int}$ marks the sum of all 2PI vacuum graphs with the dressed propagator~\eq{chi-G-LO}. It reads 
\begin{equation}
	V_\text{2PI-int} = 	\tikz [baseline=0em] {
		\draw[fill=none, line width=0.05em] (8.3em,1.0em) circle (0.6em and 0.6em);
		\draw[fill=none, line width=0.05em] (8.3em,-.2em) circle (0.6em and 0.6em);
		\draw[fill=black] (8.3em,.4em) circle (0.15em and 0.15em);
		\draw[fill=none,  line width=0.05em] (8.3em+.15em,-0.8em+.15em) -- (8.3em-.15em,-0.8em-.15em);
		\draw[fill=none,  line width=0.05em] (8.3em+.15em,-0.8em-.15em) -- (8.3em-.15em,-0.8em+.15em);
		\draw[fill=none,  line width=0.05em] (8.3em+.15em,1.6em+.15em) -- (8.3em-.15em,1.6em-.15em);
		\draw[fill=none,  line width=0.05em] (8.3em+.15em,1.6em-.15em) -- (8.3em-.15em,1.6em+.15em);
	}  \, + \, \tikz [baseline=0em] {
		\draw[fill=none, line width=0.05em] (7.2em,.40em) circle (0.6em and 0.6em);
		\draw[fill=none, line width=0.05em] (8.3em,1.0em) circle (0.6em and 0.6em);
		\draw[fill=none, line width=0.05em] (8.3em,-.2em) circle (0.6em and 0.6em);
		\draw[fill=black] (7.92em,.4em) circle (0.15em and 0.15em);
		\draw[fill=none,  line width=0.05em] (6.6em+.15em,0.4em+.15em) -- (6.6em-.15em,0.4em-.15em);
		\draw[fill=none,  line width=0.05em] (6.6em+.15em,0.4em-.15em) -- (6.6em-.15em,0.4em+.15em);
		\draw[fill=none,  line width=0.05em] (8.3em+.15em,1.6em+.15em) -- (8.3em-.15em,1.6em-.15em);
		\draw[fill=none,  line width=0.05em] (8.3em+.15em,1.6em-.15em) -- (8.3em-.15em,1.6em+.15em);
		\draw[fill=none,  line width=0.05em] (8.3em+.15em,-0.8em+.15em) -- (8.3em-.15em,-0.8em-.15em);
		\draw[fill=none,  line width=0.05em] (8.3em+.15em,-0.8em-.15em) -- (8.3em-.15em,-0.8em+.15em);
	} = \frac{N^2 \eta \varphi^2 M^2}{3840\pi^2} - \frac{N^3 \eta M^3}{46080 \pi^3} \,.
\end{equation} 
Overall, we find 
\begin{equation}\label{eq:V_2PI}
	V_\text{2PI}(\varphi,M) = \frac{\eta}{6!} \left(\varphi^2 - \frac{N M}{4\pi}\right)^3 +
	 \frac{N M^3}{24\pi}
\end{equation}
at leading order in large-$N$. Note \eq{V_2PI} assumes $ M  \geq 0$ as $M$ should be taken as the absolute value of the mass, since it stems from tadpole integrals
\begin{equation}\label{eq:tadpole-int}
	\int \!\!\!\frac{\mathrm{d}^3 p}{(2\pi)^3} \frac{1}{p^2 - M^2} = \frac{i |M|}{4\pi}\,.
\end{equation}

\begin{figure}
	\begin{tabular}{l}
		\includegraphics[scale=.45]{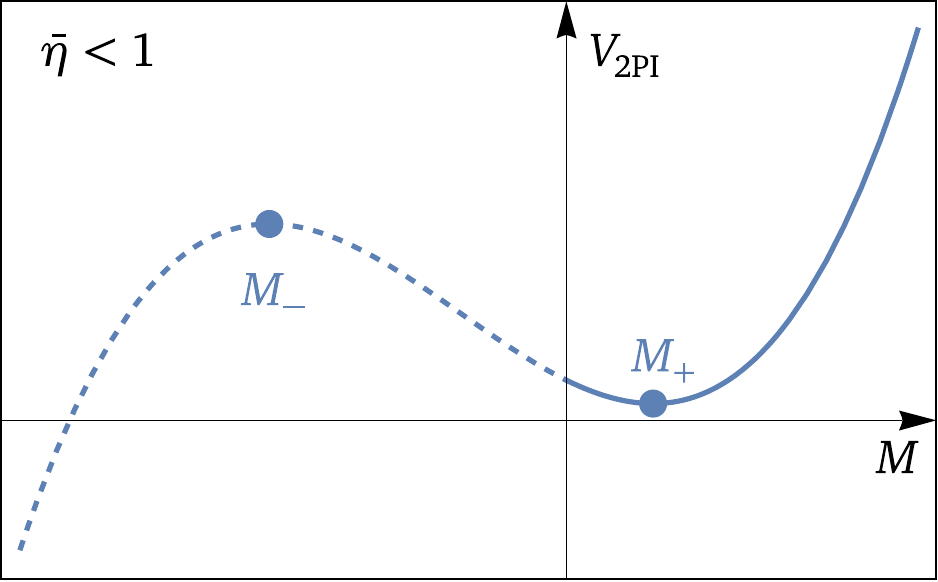}\\
		\includegraphics[scale=.45]{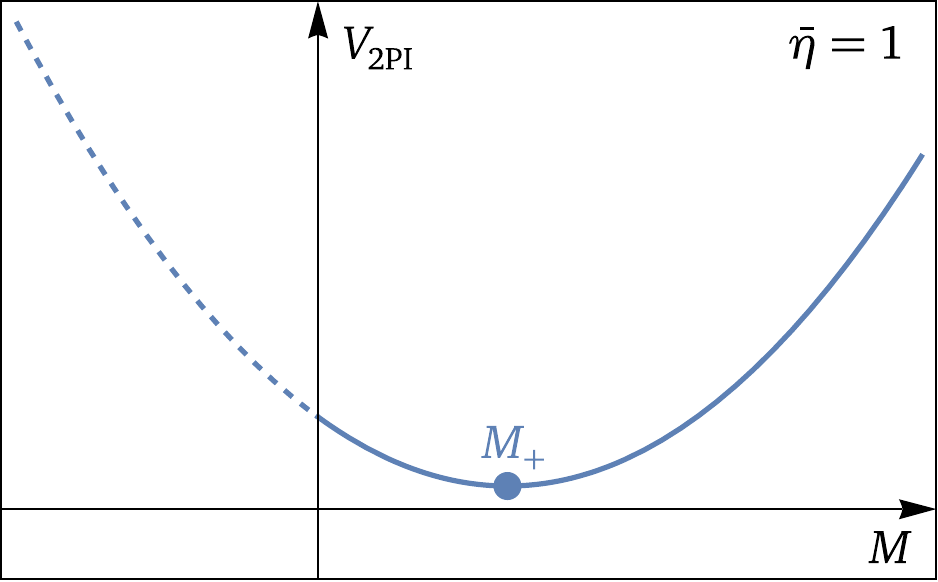}\\
		\includegraphics[scale=.45]{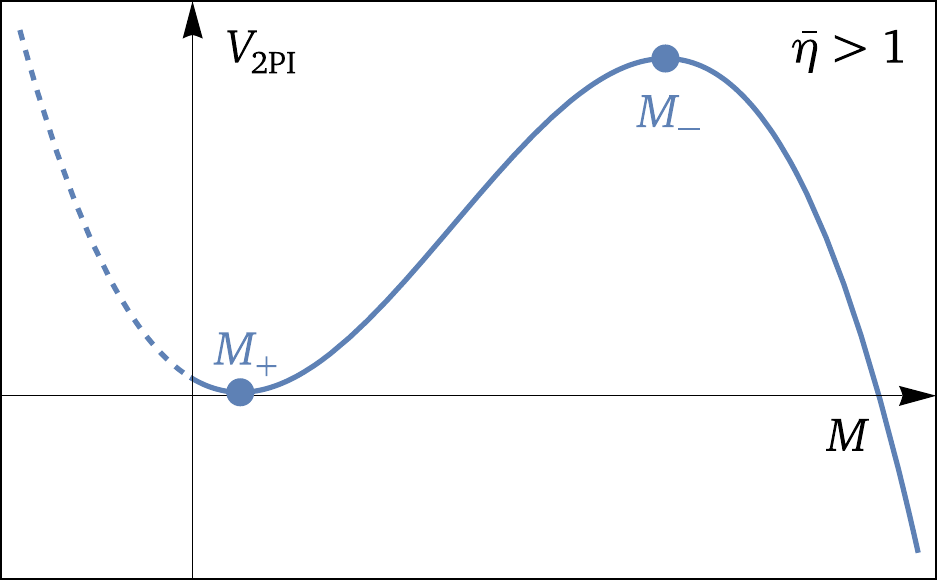}
	\end{tabular}
	\caption{$V_\text{2PI}(M)$ for a fixed value $\varphi > 0$ and $\bar{\eta} < 1$ (top panel), $\bar{\eta} = 1$ (middle panel) and $\bar{\eta} > 1$ (bottom panel).
	Only values $M \geq 0$ (solid blue line) are valid arguments of $V_\text{2PI}(M)$, while $V_\text{2PI}(M < 0)$ (dashed blue line) is only drawn for convenience.
	The extrema $M_{\pm}$~\eq{dynamical-mass} are indicated by blue dots. The minimum $M_+$ always exists, while the maximum $M_-$ is only found for $\bar{\eta} > 1$. At $\varphi = 0$, both points are shrunk together at $M_{\pm} = 0$, and $V_\text{2PI}$ becomes flat for $\bar{\eta} = 1$. The plot is adapted from \cite{Matsubara:1987iz}. }
	\label{fig:V2PI-M}
\end{figure}

The result \eq{V_2PI} has also been obtained using a variational method in the literature~\cite{Bardeen:1983rv,David:1984we,David:1985zz}.  
The 1PI effective potential is retrieved at stationary points in $M$, i.e.
\begin{equation}
	\frac{\partial V_\text{2PI}}{\partial M} = 0 \,,
\end{equation}
which yields the gap equation~\eq{gap-mass-LO}, and we recover the solutions $M_\pm$~\eq{dynamical-mass}. They are extrema of $V_\text{2PI}$ as depicted in \fig{V2PI-M}.
The perturbative solution $M_+$ indeed corresponds to a stable local minimum of the effective potential. The other solution $M_-$ is invalid for $\bar{\eta} < 1$ as $M_- < 0$, and 
has a singularity for $\bar{\eta} = 1$. For $\bar{\eta} > 1$, $M_-$ is a maximum of $V_\text{2PI}(M)$, while $M_+$ is a minimum~\cite{Matsubara:1987iz}. Thus, $M_+$ and $M_-$ represent distinct vacua of the effective potential, though only $M_+$ corresponds to the ground state as the effective potential is deeper than for $M_-$.
However, both solutions $M_\pm$ become degenerate at $\varphi = 0$. 

Inserting $M_+$ into \eq{V_2PI} yields the LO 1PI potential 
\begin{equation}\label{eq:Veff-LO}
	V_\text{eff} = \frac{(4\pi)^2 \varphi^6}{N^2}  \frac{\bar{\eta}}{6(1+\sqrt{\bar{\eta}})^2} + \mathcal{O}(N^{-3})\,.
\end{equation}
The effective potential typically contains higher-dimensional operators, which are summed up as logarithms $\log \varphi^2 / \mu$, where $\mu$ is the renormalisation scale. 
Such contributions are subleading in the large-$N$ expansion and hence absent in \eq{Veff-LO}. This can be readily understood since the effective potential is overall RG invariant, 
so the explicit RG dependence in terms $\log \varphi^2 / \mu$ needs to be cancelled by the running of $\bar{\eta}$ and $\varphi$, which are both subleading.
Furthermore, \eq{Veff-LO} na\"ively implies that the effective potential is always stable for any value of the coupling $\bar{\eta}$. 
In particular, that would render the UV fixed point \eq{Pisarski-bar} valid. However, it is commonly believed that the non-perturbative BMB phenomenon impedes this conclusion. 
We will turn to this phenomenon in a later section. Next, we extend the effective potential to order $\propto 1/N^3$.

\subsection{Next-to-leading Order Potential}\label{subsec:nlo}
We are now in a position to compute the next-to-leading order (NLO), i.e. $1/N^3$ corrections to the effective potential. To this end, we must also include interactions of the Higgs mode $H$ with the potential~\eq{V_G}
\begin{equation}
	\begin{aligned}
		V &= V_G + \frac{1}{2} \frac{\eta \varphi^4}{24} H^2 + \frac{1}{3!} \frac{\eta \varphi^3}{6} H^3 + \frac{1}{4!}\frac{\eta\varphi^2}{2} H^4  \\
		&\phantom{=\ }+ \frac{1}{5!} \eta \varphi H^5 + \frac{1}{6!} \eta H^6 + \frac{1}{2} \frac{\eta \varphi^3}{30} G^2 H \\
		&\phantom{=\ }+ \frac{1}{4} \frac{\eta \varphi^2}{10} G^2 H^2 + \frac{1}{12} \frac{\eta \varphi}{5} G^2 H^3  + \frac{1}{48} \frac{\eta}{5} G^2 H^4 \\
		&\phantom{=\ } + \frac{1}{4!} \frac{\eta \varphi}{5} G^4 H + \frac{1}{48} \frac{\eta }{5} G^4 H^2 \,,
	\end{aligned}
\end{equation}
where $G^2 \equiv G_i G_i$, and the tadpole term $\propto H$ was removed by the source term in the effective potential. In particular, the Higgs mode exhibits a mass $\tilde{m}^2 = \eta \varphi^4/24$ and gives rise to multiple interactions $\propto \eta$.

We shall now lay out a strategy to compute the effective potential at NLO. At LO, we adopted a practical version of the 2PI formalism~\cite{Cornwall:1974vz} to properly resum tadpole graphs into a mass parameter, cf. \eq{gap-LO}. This was possible because all 
LO contributions factorise. At NLO, this is not the case: dressed propagators for both $G_i$ and $H$ include loop corrections that depend explicitly on the momentum routed through the propagator.
As a result, dynamical mass parameters cannot resum all NLO corrections. Therefore, 
it is not possible to define $V_\text{2PI}$ in terms of dynamical mass parameters alone at NLO.
Nevertheless, we will retain dynamical mass parameters in order to resum all tadpole-type contributions. Other contributions to $V_\text{eff}$ are computed explicitly. 
In this sense, we employ propagators $\chi^G_{ij}$ and $\chi^H$ for the fields $G_i$ and $H$ that are \textit{half-dressed} by dynamical mass parameters $M$ and $\tilde{M}$, respectively.
Explicitly, for $\chi^G_{ij}$, we use the ansatz~\eq{chi-G-LO} while the Higgs propagator reads
\begin{equation}\label{eq:chi-H}
	\chi^H(x,y) = \braket{T H(x) H(y)} =  \int \frac{\mathrm{d}^{d}p}{(2\pi)^{d}} \frac{e^{-ip(x-y)}}{p^2 - \tilde{M}^2} \,.
\end{equation} 

A careful analysis of all possible Feynman graphs reveals that all LO and NLO contributions to the effective potential can be decomposed as 
\begin{equation}\label{eq:Veff-NLO-ansatz}
	V^\text{NLO}_\text{eff} = \frac{\eta \varphi^6}{6!}  + V_\text{1L} + V_\text{tad} + V_\text{ring} + V_\text{ct}\,.
\end{equation}
The first term is the classical potential, while $V_\text{1L,tad,ring,ct}$ are all quantum corrections from integrating out both Goldstone and Higgs modes. The $V_\text{1L}$ and $V_\text{tad}$ contain all one-loop and 2PI tadpole diagrams, yielding both contributions at LO and NLO. $V_\text{ring}$ contains additional diagrams that are purely NLO, while $V_\text{ct}$ are LO diagrams with counterterm insertions that render them NLO. We now compute \eq{Veff-NLO-ansatz} piece by piece.

$V_\text{1L}$ is obtained by applying \eq{V2PI-1L} for both Goldstone and Higgs mode, yielding
\begin{equation}
	\begin{aligned}
		V_\text{1L} &= - \frac{(N-1) M^3}{12\pi} - \frac{(N-1)M}{8\pi}\left(\frac{\eta \varphi^4}{120} - M^2\right) \\
		&\phantom{= \ } - \frac{\tilde{M}^3}{12\pi} - \frac{\tilde{M}}{8\pi}\left(\frac{\eta \varphi^4}{24} - \tilde{M}^2\right)\,.
	\end{aligned}
\end{equation}
The tadpole contributions are 
\begin{equation}
	\begin{aligned}
	V_\text{tad} &= \tikz [baseline=0em] {
		\draw[fill=none, line width=0.05em] (8.3em,1.0em) circle (0.6em and 0.6em);
		\draw[fill=none, line width=0.05em] (8.3em,-.2em) circle (0.6em and 0.6em);
		\draw[fill=black] (8.3em,.4em) circle (0.15em and 0.15em);
		\draw[fill=none,  line width=0.05em] (8.3em+.15em,-0.8em+.15em) -- (8.3em-.15em,-0.8em-.15em);
		\draw[fill=none,  line width=0.05em] (8.3em+.15em,-0.8em-.15em) -- (8.3em-.15em,-0.8em+.15em);
		\draw[fill=none,  line width=0.05em] (8.3em+.15em,1.6em+.15em) -- (8.3em-.15em,1.6em-.15em);
		\draw[fill=none,  line width=0.05em] (8.3em+.15em,1.6em-.15em) -- (8.3em-.15em,1.6em+.15em);
	}  \, + \, \tikz [baseline=0em] {
		\draw[fill=none, line width=0.05em] (7.2em,.40em) circle (0.6em and 0.6em);
		\draw[fill=none, line width=0.05em] (8.3em,1.0em) circle (0.6em and 0.6em);
		\draw[fill=none, line width=0.05em] (8.3em,-.2em) circle (0.6em and 0.6em);
		\draw[fill=black] (7.92em,.4em) circle (0.15em and 0.15em);
		\draw[fill=none,  line width=0.05em] (6.6em+.15em,0.4em+.15em) -- (6.6em-.15em,0.4em-.15em);
		\draw[fill=none,  line width=0.05em] (6.6em+.15em,0.4em-.15em) -- (6.6em-.15em,0.4em+.15em);
		\draw[fill=none,  line width=0.05em] (8.3em+.15em,1.6em+.15em) -- (8.3em-.15em,1.6em-.15em);
		\draw[fill=none,  line width=0.05em] (8.3em+.15em,1.6em-.15em) -- (8.3em-.15em,1.6em+.15em);
		\draw[fill=none,  line width=0.05em] (8.3em+.15em,-0.8em+.15em) -- (8.3em-.15em,-0.8em-.15em);
		\draw[fill=none,  line width=0.05em] (8.3em+.15em,-0.8em-.15em) -- (8.3em-.15em,-0.8em+.15em);
	} \, + \, \tikz [baseline=0em] {
		\draw[fill=none, line width=0.05em] (8.3em,1.0em) circle (0.6em and 0.6em);
		\draw[fill=none,  dash pattern=on .1em off .1em, line width=0.05em] (8.3em,-.2em) circle (0.6em and 0.6em);
		\draw[fill=black] (8.3em,.4em) circle (0.15em and 0.15em);
		\draw[fill=none,  line width=0.05em] (8.3em+.15em,-0.8em+.15em) -- (8.3em-.15em,-0.8em-.15em);
		\draw[fill=none,  line width=0.05em] (8.3em+.15em,-0.8em-.15em) -- (8.3em-.15em,-0.8em+.15em);
		\draw[fill=none,  line width=0.05em] (8.3em+.15em,1.6em+.15em) -- (8.3em-.15em,1.6em-.15em);
		\draw[fill=none,  line width=0.05em] (8.3em+.15em,1.6em-.15em) -- (8.3em-.15em,1.6em+.15em);
	}  \, + \, \tikz [baseline=0em] {
		\draw[fill=none, line width=0.05em] (7.2em,.40em) circle (0.6em and 0.6em);
		\draw[fill=none, line width=0.05em] (8.3em,1.0em) circle (0.6em and 0.6em);
		\draw[fill=none, dash pattern=on .1em off .1em, line width=0.05em] (8.3em,-.2em) circle (0.6em and 0.6em);
		\draw[fill=black] (7.92em,.4em) circle (0.15em and 0.15em);
		\draw[fill=none,  line width=0.05em] (6.6em+.15em,0.4em+.15em) -- (6.6em-.15em,0.4em-.15em);
		\draw[fill=none,  line width=0.05em] (6.6em+.15em,0.4em-.15em) -- (6.6em-.15em,0.4em+.15em);
		\draw[fill=none,  line width=0.05em] (8.3em+.15em,1.6em+.15em) -- (8.3em-.15em,1.6em-.15em);
		\draw[fill=none,  line width=0.05em] (8.3em+.15em,1.6em-.15em) -- (8.3em-.15em,1.6em+.15em);
		\draw[fill=none,  line width=0.05em] (8.3em+.15em,-0.8em+.15em) -- (8.3em-.15em,-0.8em-.15em);
		\draw[fill=none,  line width=0.05em] (8.3em+.15em,-0.8em-.15em) -- (8.3em-.15em,-0.8em+.15em);
	} \\
	&= \frac{\bar{\eta} }2 (\varphi M)^2  - \frac{N+3}{24\pi} \bar{\eta}M^3 + \frac{3\bar{\eta}}{N}  \varphi^2 \tilde{M} M  
	- \frac{\bar{\eta}}{8\pi} \tilde{M} M^2 
	\end{aligned}
\end{equation}
when expanded up to NLO. Here, the dashed lines \begin{tikzpicture}[baseline=-.3em]
	\draw[fill=none, dash pattern=on .1em off .1em, line width=0.05em] (-1.0em,0) -- (.0em,0);
\end{tikzpicture} represent propagators of the Higgs, the solid lines correspond to the Goldstone mode, and crosses denote dressed propagators in either case.
This leads to the intermediate result
\begin{equation}
	\begin{aligned}
		V_\text{1L} + V_\text{tad} &= \frac{(N-1) M^3}{24\pi} + \frac{8 \pi^2 \bar{\eta}}{3N^2} \left(\varphi^2 - \frac{M N}{4\pi}\right)^3  \\
		&\phantom{=\ } + \frac{\tilde{M}^3 }{24\pi}  - \frac{2\pi\bar{\eta}}{N^2} \left(\varphi^2 - \frac{M N}{4\pi}\right)^2 (\tilde{M} + M)  \\
		&\phantom{=\ } + \frac{4 \pi \bar{\eta}}{N^2} \left(\varphi^2 - \frac{M N}{4\pi}\right)  \varphi^2 (M - 2 \tilde{M})\,.\\
	\end{aligned}
\end{equation}
Tadpoles are resummed when inserting the $\tilde{M}$ at the stationary value
\begin{equation}\label{eq:2PI-stationary}
	\frac{\partial (V_\text{1L} + V_\text{tad}) }{\partial \tilde{M}} = 0
\end{equation}
which yields the gap equation
\begin{equation}\label{eq:gap-Higgs}
	\begin{tikzpicture}[baseline=-.3em]
			\draw[fill=none, dash pattern=on .1em off .1em, line width=0.05em] (-1.0em,0) -- (1em,0);
			\draw[fill=none, line width=0.05em] (-.15em,-0.15em) -- (+.15em,0.15em);
			\draw[fill=none, line width=0.05em] (+.15em,-0.15em) -- (-.15em,0.15em);
		\end{tikzpicture} = \ \begin{tikzpicture}[baseline=-.3em]
			\draw[fill=none, dash pattern=on .1em off .1em, line width=0.05em] (-1.0em,0) -- (1em,0);
		\end{tikzpicture} + \begin{tikzpicture}[baseline=-.3em]
			\draw[fill=none, dash pattern=on .1em off .1em, line width=0.05em] (-1.5em,0) -- (0,0);
			\draw[fill=none, dash pattern=on .1em off .1em, line width=0.05em] (+1.5em,0) -- (0,0);
			\draw[fill=none, line width=0.05em] (0em,+.6em) circle (0.6em and 0.6em);
			\draw[fill=none, line width=0.05em] (-.15em,1.35em) -- (+.15em,1.05em);
			\draw[fill=none, line width=0.05em] (+.15em,1.35em) -- (-.15em,1.05em);
			\draw[fill=black] (0,0) circle (0.15em and 0.15em);
			\draw[fill=none, line width=0.05em] (+.8em,-0.15em) -- (+1.1em,+0.15em);
			\draw[fill=none, line width=0.05em] (+.8em,+0.15em) -- (+1.1em,-0.15em);
		\end{tikzpicture}  + \begin{tikzpicture}[baseline=-.3em]
			\draw[fill=none, dash pattern=on .1em off .1em, line width=0.05em] (-1.5em,0) -- (0,0);
			\draw[fill=none, dash pattern=on .1em off .1em, line width=0.05em] (+1.5em,0) -- (0,0);
			\draw[fill=none, line width=0.05em] (0em,-.6em) circle (0.6em and 0.6em);
			\draw[fill=none, line width=0.05em] (0em,+.6em) circle (0.6em and 0.6em);
			\draw[fill=none, line width=0.05em] (-.15em,1.35em) -- (+.15em,1.05em);
			\draw[fill=none, line width=0.05em] (+.15em,1.35em) -- (-.15em,1.05em);
			\draw[fill=none, line width=0.05em] (-.15em,-1.35em) -- (+.15em,-1.05em);
			\draw[fill=none, line width=0.05em] (+.15em,-1.35em) -- (-.15em,-1.05em);
			\draw[fill=black] (0,0) circle (0.15em and 0.15em);
			\draw[fill=none, line width=0.05em] (+.8em,-0.15em) -- (+1.1em,+0.15em);
			\draw[fill=none, line width=0.05em] (+.8em,+0.15em) -- (+1.1em,-0.15em);
		\end{tikzpicture}\quad .
	\end{equation}
Computing \eq{gap-Higgs} explicitly, we obtain
\begin{equation}
 \tilde{M}^2 = \frac{\eta}{24} \varphi^4 - \frac{N \eta M \varphi^2}{80\pi} + \frac{N^2 \eta M^2}{1920 \pi^2}\,.
\end{equation}
Inserting the leading-order mass $M=M_+$~\eq{dynamical-mass}, this simplifies to
\begin{equation}\label{eq:Mtilde}
	\tilde{M} = \frac{4\pi \varphi^2}{N} \frac{\sqrt{ 5 + 4\sqrt{\bar{\eta}} }}{1 + \bar{\eta}^{-1/2}}\,.
\end{equation}
Similarly, computing the stationary value of $M$ via the condition
\begin{equation}
	\frac{\partial (V_\text{1L} + V_\text{tad}) }{\partial M} = 0
\end{equation}
yields the NLO corrections $\delta M$ to Goldstone mass, i.e. $M = M_+ + \delta M$. 
However, the value of $\delta M$ does not contribute to the NLO effective potential as such a term would read 
\begin{equation}
	\delta M \frac{\partial (V_\text{1L} + V_\text{tad}) }{\partial M} \bigg|_{M=M_+}  = 0\,.
\end{equation}

Now we consider the correction $V_\text{ring}$ in \eq{Veff-NLO-ansatz}. It is a sum of two families of ring diagrams, i.e.
\begin{equation}\label{eq:V_ring}
	V_\text{ring} = \sum_{n=2}^\infty R_n + \sum_{n=1}^\infty \tilde{R}_{n}\,.
\end{equation}
The ring diagrams $R_n$ and $\tilde{R}_{n}$ each consist of subgraphs of the sort 
\begin{equation}\label{eq:dressed-bubble}
	\tikz [baseline=-0.3em] {
		\draw[fill=black] (+0.8em,0em) circle (0.15em and 0.15em);
		\draw[fill=black] (-0.8em,0em) circle (0.15em and 0.15em);
		\draw[fill=none, pattern=north east lines, line width=0.05em] (0em,0em) circle (0.8em and 0.8em);
	} \quad = \quad  \tikz [baseline=-0.3em] {

		\draw[fill=black] (+0.8em,0em) circle (0.15em and 0.15em);
		\draw[fill=black] (-0.8em,0em) circle (0.15em and 0.15em);
		\draw[fill=none,  line width=0.05em] (0em+.15em,.8em+.15em) -- (0em-.15em,.8em-.15em);
		\draw[fill=none,  line width=0.05em] (0em+.15em,.8em-.15em) -- (0em-.15em,.8em+.15em);
		\draw[fill=none,  line width=0.05em] (0em+.15em,-.8em+.15em) -- (0em-.15em,-.8em-.15em);
		\draw[fill=none,  line width=0.05em] (0em+.15em,-.8em-.15em) -- (0em-.15em,-.8em+.15em);
		\draw[fill=none, line width=0.05em] (0em,0em) circle (0.8em and 0.8em);	
	} \quad + \quad \tikz [baseline=-0.3em] {
		\draw[fill=black] (+0.8em,0em) circle (0.15em and 0.15em);
		\draw[fill=black] (-0.8em,0em) circle (0.15em and 0.15em);
		\draw[fill=none, line width=0.05em] (0em,0em) circle (0.70em and 0.40em);
		\draw[fill=none, line width=0.05em] (0.8em,0.75em) circle (0.3em and 0.6em);
		\draw[fill=none,  line width=0.05em] (0em+.15em,.4em+.15em) -- (0em-.15em,.4em-.15em);
		\draw[fill=none,  line width=0.05em] (0em+.15em,.4em-.15em) -- (0em-.15em,.4em+.15em);	
		\draw[fill=none,  line width=0.05em] (0em+.15em,-.4em+.15em) -- (0em-.15em,-.4em-.15em);
		\draw[fill=none,  line width=0.05em] (0em+.15em,-.4em-.15em) -- (0em-.15em,-.4em+.15em);
		\draw[fill=none,  line width=0.05em] (1.1em+.15em,0.7em+.15em) -- (1.1em-.15em,0.7em-.15em);
		\draw[fill=none,  line width=0.05em] (1.1em+.15em,0.7em-.15em) -- (1.1em-.15em,0.7em+.15em);	
	}\,.
\end{equation} 
The first family, $R_n$, only involves the Goldstone and consists of $n$ pieces
\begin{equation}\label{eq:Rn}
	R_n = \quad \tikz [baseline=-1.2em] {
		\draw[fill=black] (+0.8em,0em) circle (0.15em and 0.15em);
		\draw[fill=black] (-0.8em,0em) circle (0.15em and 0.15em);
		\draw[fill=none, pattern=north east lines, line width=0.05em] (0em,0em) circle (0.8em and 0.8em);
		\draw[pattern=north east lines] (-0.8em,0em) arc (90:270:1.2em) -- (+0.8em,-2.4em) arc (-450:-270:1.2em) arc  (-270:-450:0.8em) --  (-0.8em,-1.6em) arc (270:90:0.8em) ;
		\node[anchor=east] at (-0.1em,0em) {$\bigg[ $};
		\node[anchor=west] at (+0.1em,0em) {$\bigg]^{n-1}$};
	} 
\end{equation} 
such that the connection between each piece is a quartic or sextic vertex.
For instance,   
\begin{equation}
	R_{2,3,4,5,\dots} = \quad \tikz[baseline=-.4em] {
		\draw[pattern=north east lines] (0em,0em) to[bend left=45] (1.2em,1.2em) to [bend left=45] (2.4em,0em) to [bend right=15] (1.2em,.4em) to [bend right=15] (0em,0em);
		\draw[pattern=north east lines] (0em,0em) to[bend right=45] (1.2em,-1.2em) to [bend right=45] (2.4em,0em) to [bend left=15] (1.2em,-.4em) to [bend left=15] (0em,0em);
		\draw[fill=black] (0em,0em) circle (0.15em and 0.15em);
		\draw[fill=black] (2.4em,0em) circle (0.15em and 0.15em);
	}\,, \ \
	\tikz[baseline=0em] {
		\draw[pattern=north east lines] (0em,0em) to[bend right=30] (2.em,0em) to[bend right=30] (0em,0em);
		\draw[pattern=north east lines] (0em,0em) to[bend left=30] (1.em,1.73em) to[bend left=30] (0em,0em);
		\draw[pattern=north east lines] (2.0em,0em) to[bend right=30] (1.em,1.73em) to[bend right=30] (2.em,0em);
		\draw[fill=black] (0em,0em) circle (0.15em and 0.15em);
		\draw[fill=black] (2.em,0em) circle (0.15em and 0.15em);
		\draw[fill=black] (1.em,1.73em) circle (0.15em and 0.15em);
	}\,, \ \ 
	\tikz[baseline=0em] {
		\draw[pattern=north east lines] (0em,0em) to[bend right=45] (1.6em,0em) to[bend right=45] (0em,0em);
		\draw[pattern=north east lines] (0em,0em) to[bend left=45] (0em,1.6em) to[bend left=45] (0em,0em);
		\draw[pattern=north east lines] (1.6em,1.6em) to[bend left=45] (1.6em,0em) to[bend left=45] (1.6em,1.6em);
		\draw[pattern=north east lines] (1.6em,1.6em) to[bend right=45] (0em,1.6em) to[bend right=45] (1.6em,1.6em);
		\draw[fill=black] (0.em,0.em) circle (0.15em and 0.15em);
		\draw[fill=black] (1.6em,0.em) circle (0.15em and 0.15em);
		\draw[fill=black] (0.em,1.6em) circle (0.15em and 0.15em);
		\draw[fill=black] (1.6em,1.6em) circle (0.15em and 0.15em);
	},\ \ 
	\tikz[baseline=-.8em] {
		\draw[pattern=north east lines] (1.2*0.851em,1.2*0.000em)  to[bend right=45] (1.2*0.263em,1.2*0.809em) to[bend right=45] (1.2*0.851em,1.2*0.000em);
		\draw[pattern=north east lines] (1.2*0.263em,1.2*0.809em)  to[bend right=45] (-1.2*0.688em,1.2*0.500em) to[bend right=45] (1.2*0.263em,1.2*0.809em);
		\draw[pattern=north east lines] (-1.2*0.688em,1.2*0.500em)  to[bend right=45] (-1.2*0.688em,-1.2*0.500em) to[bend right=45] (-1.2*0.688em,1.2*0.500em);
		\draw[pattern=north east lines] (-1.2*0.688em,-1.2*0.500em)  to[bend right=45] (1.2*0.263em,-1.2*0.809em) to[bend right=45] (-1.2*0.688em,-1.2*0.500em);
		\draw[pattern=north east lines] (1.2*0.263em,-1.2*0.809em)  to[bend right=45] (1.2*0.851em,1.2*0.000em) to[bend right=45] (1.2*0.263em,-1.2*0.809em);

		\draw[fill=black] (1.2*0.851em,1.2*0.000em) circle (0.15em and 0.15em);
		\draw[fill=black] (1.2*0.263em,1.2*0.809em) circle (0.15em and 0.15em);
		\draw[fill=black] (-1.2*0.688em,1.2*0.500em) circle (0.15em and 0.15em);
		\draw[fill=black] (-1.2*0.688em,-1.2*0.500em) circle (0.15em and 0.15em);
		\draw[fill=black] (1.2*0.263em,-1.2*0.809em) circle (0.15em and 0.15em);
	},\,\dots \,.
\end{equation}
We relegate the details regarding the computations to App.~\ref{sec:appendix} and present the results here.
The first two diagrams $R_2$ and $R_3$ are UV divergent. 
We obtain  
\begin{equation}\label{eq:R2}
	\begin{aligned}
		R_2 &= \frac{4\pi \bar{\eta}^2 M \left(\varphi^2 - \frac{N M}{4\pi} \right)^2 }{N^2 \varepsilon}  \\
		&\phantom{= \ }- \frac{4\bar{\eta}^2 M^2 \left(\varphi^2 - \frac{N M}{4\pi} \right)}{N}\left[1 - \log 2 - \log\frac{M}{\mu}\right] \\
		 &\phantom{= \ } + \frac{8\pi \bar{\eta}^2 M \left(\varphi^2 - \frac{N M}{4\pi} \right)^2}{N^2}\left[4 - 5 \log 2 - 3 \log \frac{M}{\mu}\right]
	\end{aligned}
\end{equation}
and 
\begin{equation}\label{eq:R3}
	\begin{aligned}
		R_3 &= \frac{2\pi^{4} \bar{\eta}^3 \left(\varphi^2 - \frac{N M}{4\pi} \right)^3}{3 N^3 \varepsilon} \\ 
		&\phantom{= \ } + \frac{4\pi^{4} \bar{\eta}^3 \left(\varphi^2 - \frac{N M}{4\pi} \right)^3}{3 N^3}\left[1 - \frac{ 42 \zeta_3}{\pi^2} - 2 \log \frac{2M^2}{\mu^2}\right]\\
		&\phantom{= \ }- \frac{\pi^{3} \bar{\eta}^3 M \left(\varphi^2 - \frac{N M}{4\pi} \right)^2}{ N^2} \left[1 - \log 2 - \log \frac{M}{\mu}\right]\,.
	\end{aligned}
\end{equation}
For $n \geq 4$, the $R_n$ are finite and read
\begin{equation}\label{eq:Rn_sol}
	R_n = - \frac{2 M^3}{\pi^2 n } \left[\bar{\eta} \left(1- \frac{4\pi \varphi^2}{N M} \right)\right]^n \int_0^\infty \!\!\!  \frac{\arctan^n(z)}{z^{n-2}}\mathrm{d} z\,. 
\end{equation}
Diagram summation and momentum integration may be exchanged if the results are convergent, and we obtain 
\begin{equation}\label{eq:RnLarger4}
	\sum_{n=4}^\infty R_n = \frac{2M^3}{\pi^2} f\left(\bar{\eta} \left(1-\frac{4\pi \varphi^2}{N M} \right)\right)
\end{equation} with the function 
\begin{equation}\label{eq:ff1}
	\begin{aligned}
		f(x) &= \int_{0}^\infty \mathrm{d}z \bigg[ x z \arctan(z)  + \frac{1}{2} x^2 \arctan^2 z\\	
		&\qquad   \frac{x^3 \arctan^3 z }{3z}  +  z^2 \log\left(1 - \frac{x \arctan z}{z}\right) \bigg]\,,
	\end{aligned}
\end{equation}
which can be evaluated numerically within its radius of convergence $x \leq 1$.
The second family $\tilde{R}_n$ consists of several sequences of subgraphs~\eq{dressed-bubble}, i.e.
\begin{equation}
	\begin{aligned}
	\tikz [baseline=-0.3em] {
		\draw[fill=black] (+0.8em,0em) circle (0.15em and 0.15em);
		\draw[fill=black] (-0.8em,0em) circle (0.15em and 0.15em);
		\draw[fill=none, pattern=crosshatch, line width=0.05em] (0em,0em) circle (0.8em and 0.8em);
	} \ &= \   \sum_{k=1}^\infty \left(\tikz [baseline=-0.3em] {
		\draw[fill=black] (+0.8em,0em) circle (0.15em and 0.15em);
		\draw[fill=black] (-0.8em,0em) circle (0.15em and 0.15em);
		\draw[fill=none, pattern=north east lines, line width=0.05em] (0em,0em) circle (0.8em and 0.8em);
		\node[anchor=east] at (-0.1em,0em) {$\bigg[ $};
		\node[anchor=west] at (+0.1em,0em) {$\bigg]^k$};
	}   +  \tikz [baseline=-0.3em] {
		\draw[fill=black] (+0.8em,0em) circle (0.15em and 0.15em);
		\draw[fill=black] (-0.8em,0em) circle (0.15em and 0.15em);
		\draw[fill=none, pattern=north east lines, line width=0.05em] (0em,0em) circle (0.8em and 0.8em);
		\draw[fill=none, line width=0.05em, rotate around={45:(-0.95em,0.15em)}] (-0.95em,.55em) circle (0.2em and 0.5em);
		\draw[fill=none,  line width=0.05em] (-1.6em+0.15em,0.8em+0.15em) -- (-1.6em-0.15em,0.8em-0.15em);
		\draw[fill=none,  line width=0.05em] (-1.6em+0.15em,0.8em-0.15em) -- (-1.6em-0.15em,0.8em+0.15em);
		\node[anchor=east] at (-0.1em,0em) {$\bigg[ $};
		\node[anchor=west] at (+0.1em,0em) {$\bigg]^k$};
	} \right)\\
	&= \tikz [baseline=-0.3em,scale=0.9] {
		\draw[fill=black] (+0.8em,0em) circle (0.15em and 0.15em);
		\draw[fill=black] (-0.8em,0em) circle (0.15em and 0.15em);
		\draw[fill=none, line width=0.05em] (0em,0em) circle (0.70em and 0.40em);
		\draw[fill=none,  line width=0.05em] (0em+.15em,.4em+.15em) -- (0em-.15em,.4em-.15em);
		\draw[fill=none,  line width=0.05em] (0em+.15em,.4em-.15em) -- (0em-.15em,.4em+.15em);	
		\draw[fill=none,  line width=0.05em] (0em+.15em,-.4em+.15em) -- (0em-.15em,-.4em-.15em);
		\draw[fill=none,  line width=0.05em] (0em+.15em,-.4em-.15em) -- (0em-.15em,-.4em+.15em);
	} + \tikz [baseline=-0.3em,scale=0.9] {
		\draw[fill=black] (+0.8em,0em) circle (0.15em and 0.15em);
		\draw[fill=black] (-0.8em,0em) circle (0.15em and 0.15em);
		\draw[fill=none, line width=0.05em] (0em,0em) circle (0.70em and 0.40em);
		\draw[fill=none, line width=0.05em] (-0.8em,0.75em) circle (0.3em and 0.6em);
		\draw[fill=none,  line width=0.05em] (0em+.15em,.4em+.15em) -- (0em-.15em,.4em-.15em);
		\draw[fill=none,  line width=0.05em] (0em+.15em,.4em-.15em) -- (0em-.15em,.4em+.15em);	
		\draw[fill=none,  line width=0.05em] (0em+.15em,-.4em+.15em) -- (0em-.15em,-.4em-.15em);
		\draw[fill=none,  line width=0.05em] (0em+.15em,-.4em-.15em) -- (0em-.15em,-.4em+.15em);
		\draw[fill=none,  line width=0.05em] (-1.1em+.15em,0.7em+.15em) -- (-1.1em-.15em,0.7em-.15em);
		\draw[fill=none,  line width=0.05em] (-1.1em+.15em,0.7em-.15em) -- (-1.1em-.15em,0.7em+.15em);		
	}+ \tikz [baseline=-0.3em,scale=0.9] {
		\draw[fill=black] (+0.8em,0em) circle (0.15em and 0.15em);
		\draw[fill=black] (-0.8em,0em) circle (0.15em and 0.15em);
		\draw[fill=none, line width=0.05em] (0em,0em) circle (0.70em and 0.40em);
		\draw[fill=none, line width=0.05em] (0.8em,0.75em) circle (0.3em and 0.6em);
		\draw[fill=none,  line width=0.05em] (0em+.15em,.4em+.15em) -- (0em-.15em,.4em-.15em);
		\draw[fill=none,  line width=0.05em] (0em+.15em,.4em-.15em) -- (0em-.15em,.4em+.15em);	
		\draw[fill=none,  line width=0.05em] (0em+.15em,-.4em+.15em) -- (0em-.15em,-.4em-.15em);
		\draw[fill=none,  line width=0.05em] (0em+.15em,-.4em-.15em) -- (0em-.15em,-.4em+.15em);
		\draw[fill=none,  line width=0.05em] (1.1em+.15em,0.7em+.15em) -- (1.1em-.15em,0.7em-.15em);
		\draw[fill=none,  line width=0.05em] (1.1em+.15em,0.7em-.15em) -- (1.1em-.15em,0.7em+.15em);	
	} + \tikz [baseline=-0.3em,scale=0.9] {
		\draw[fill=black] (+0.8em,0em) circle (0.15em and 0.15em);
		\draw[fill=black] (-0.8em,0em) circle (0.15em and 0.15em);
		\draw[fill=none, line width=0.05em] (0em,0em) circle (0.70em and 0.40em);
		\draw[fill=none, line width=0.05em] (-0.8em,0.75em) circle (0.3em and 0.6em);
		\draw[fill=none, line width=0.05em] (0.8em,0.75em) circle (0.3em and 0.6em);
		\draw[fill=none,  line width=0.05em] (0em+.15em,.4em+.15em) -- (0em-.15em,.4em-.15em);
		\draw[fill=none,  line width=0.05em] (0em+.15em,.4em-.15em) -- (0em-.15em,.4em+.15em);	
		\draw[fill=none,  line width=0.05em] (0em+.15em,-.4em+.15em) -- (0em-.15em,-.4em-.15em);
		\draw[fill=none,  line width=0.05em] (0em+.15em,-.4em-.15em) -- (0em-.15em,-.4em+.15em);
		\draw[fill=none,  line width=0.05em] (1.1em+.15em,0.7em+.15em) -- (1.1em-.15em,0.7em-.15em);
		\draw[fill=none,  line width=0.05em] (1.1em+.15em,0.7em-.15em) -- (1.1em-.15em,0.7em+.15em);	
		\draw[fill=none,  line width=0.05em] (-1.1em+.15em,0.7em+.15em) -- (-1.1em-.15em,0.7em-.15em);
		\draw[fill=none,  line width=0.05em] (-1.1em+.15em,0.7em-.15em) -- (-1.1em-.15em,0.7em+.15em);	
	} + \tikz [baseline=-0.3em,scale=0.9] {
		\draw[fill=black] (+0.8em,0em) circle (0.15em and 0.15em);
		\draw[fill=black] (-0.8em,0em) circle (0.15em and 0.15em);
		\draw[fill=none, line width=0.05em] (0em,0em) circle (0.70em and 0.40em);
		\draw[fill=none,  line width=0.05em] (0em+.15em,.4em+.15em) -- (0em-.15em,.4em-.15em);
		\draw[fill=none,  line width=0.05em] (0em+.15em,.4em-.15em) -- (0em-.15em,.4em+.15em);	
		\draw[fill=none,  line width=0.05em] (0em+.15em,-.4em+.15em) -- (0em-.15em,-.4em-.15em);
		\draw[fill=none,  line width=0.05em] (0em+.15em,-.4em-.15em) -- (0em-.15em,-.4em+.15em); 
		\draw[fill=black] (+2.4em,0em) circle (0.15em and 0.15em);
		\draw[fill=none, line width=0.05em] (1.6em,0em) circle (0.70em and 0.40em);
		\draw[fill=none,  line width=0.05em] (1.6em+.15em,.4em+.15em) -- (1.6em-.15em,.4em-.15em);
		\draw[fill=none,  line width=0.05em] (1.6em+.15em,.4em-.15em) -- (1.6em-.15em,.4em+.15em);	
		\draw[fill=none,  line width=0.05em] (1.6em+.15em,-.4em+.15em) -- (1.6em-.15em,-.4em-.15em);
		\draw[fill=none,  line width=0.05em] (1.6em+.15em,-.4em-.15em) -- (1.6em-.15em,-.4em+.15em);
		} + \cdots \,,
	\end{aligned}
\end{equation}
such that they are connected by $n$ lines of the Higgs field
\begin{equation}\label{eq:Rtilden-dia}
	\tilde{R}_{n} = \quad \tikz [baseline=-1.2em] {
		\draw[fill=black] (+0.8em,0em) circle (0.15em and 0.15em);
		\draw[fill=black] (-0.8em,0em) circle (0.15em and 0.15em);
		\draw[fill=none, pattern=crosshatch, line width=0.05em] (0em,0em) circle (0.8em and 0.8em);
		\draw[dash pattern=on .1em off .1em] (-0.8em,0em) arc (90:270:1.2em) -- (+0.8em,-2.4em);
		\draw[dash pattern=on .1em off .1em] (+0.8em,0em) arc (-270:-450:1.2em);
		\draw[fill=none,  line width=0.05em] (0em + 0.15em,-2.4em+0.15em) -- (0em-.15em,-2.4em-0.15em);
		\draw[fill=none,  line width=0.05em] (0em + 0.15em,-2.4em-0.15em) -- (0em-.15em,-2.4em+0.15em);
		\node[anchor=east] at (-0.5em,0em) {$\bigg[ $};
		\node[anchor=west] at (+0.5em,0em) {$\bigg]^n$};
	} \quad .
\end{equation}
In the following, we first consider the diagrams belonging to $\tilde{R}_1$, which contain a single Higgs line. We divide them into a subfamily $\tilde{R}_{1,l}$, where $l$ labels the number of insertions of the blob~\eq{dressed-bubble}
\begin{equation}
	\tilde{R}_{1,l} = \quad \tikz [baseline=-1.2em] {
		\draw[fill=black] (+0.8em,0em) circle (0.15em and 0.15em);
		\draw[fill=black] (-0.8em,0em) circle (0.15em and 0.15em);
		\draw[fill=none, pattern=north east lines, line width=0.05em] (0em,0em) circle (0.8em and 0.8em);
		\draw[dash pattern=on .1em off .1em] (-0.8em,0em) arc (90:270:1.2em) -- (+0.8em,-2.4em);
		\draw[dash pattern=on .1em off .1em] (+0.8em,0em) arc (-270:-450:1.2em);
		\draw[fill=none,  line width=0.05em] (0em + 0.15em,-2.4em+0.15em) -- (0em-.15em,-2.4em-0.15em);
		\draw[fill=none,  line width=0.05em] (0em + 0.15em,-2.4em-0.15em) -- (0em-.15em,-2.4em+0.15em);
		\node[anchor=east] at (-0.1em,0em) {$\bigg[ $};
		\node[anchor=west] at (+0.1em,0em) {$\bigg]^l$};
	} \quad  + \quad \tikz [baseline=-1.2em] {
		\draw[fill=black] (+0.8em,0em) circle (0.15em and 0.15em);
		\draw[fill=black] (-0.8em,0em) circle (0.15em and 0.15em);
		\draw[fill=none, pattern=north east lines, line width=0.05em] (0em,0em) circle (0.8em and 0.8em);
		\draw[dash pattern=on .1em off .1em] (-0.8em,0em) arc (90:270:1.2em) -- (+0.8em,-2.4em);
		\draw[dash pattern=on .1em off .1em] (+0.8em,0em) arc (-270:-450:1.2em);
		\draw[fill=none,  line width=0.05em] (0em + 0.15em,-2.4em+0.15em) -- (0em-.15em,-2.4em-0.15em);
		\draw[fill=none,  line width=0.05em] (0em + 0.15em,-2.4em-0.15em) -- (0em-.15em,-2.4em+0.15em);
		\draw[fill=none, line width=0.05em, rotate around={45:(-0.95em,0.15em)}] (-0.95em,.55em) circle (0.2em and 0.5em);
		\draw[fill=none,  line width=0.05em] (-1.6em+0.15em,0.8em+0.15em) -- (-1.6em-0.15em,0.8em-0.15em);
		\draw[fill=none,  line width=0.05em] (-1.6em+0.15em,0.8em-0.15em) -- (-1.6em-0.15em,0.8em+0.15em);
		\node[anchor=east] at (-0.1em,0em) {$\bigg[ $};
		\node[anchor=west] at (+0.1em,0em) {$\bigg]^l$};
	} \quad .
\end{equation}
Only the graphs $\tilde{R}_{1,1}$ are UV divergent
\begin{equation}\label{eq:R1}
	\begin{aligned}
		\tilde{R}_{1,1} &= \quad \tikz [baseline=0.1em] {
			\draw[fill=none,  line width=0.05em] (7.2em,.40em) circle (0.6em and 0.6em);
			\draw[fill=none, dash pattern=on .1em off .1em, line width=0.05em] (7.2em,1.0em) -- (7.2em,-.2em);
			\draw[fill=black] (7.2em,1.0em) circle (0.10em and 0.10em);
			\draw[fill=black] (7.2em,-0.2em) circle (0.10em and 0.10em);
			\draw[fill=none,  line width=0.05em] (6.6em+.15em,0.4em+.15em) -- (6.6em-.15em,0.4em-.15em);
			\draw[fill=none,  line width=0.05em] (6.6em+.15em,0.4em-.15em) -- (6.6em-.15em,0.4em+.15em);
			\draw[fill=none,  line width=0.05em] (7.8em+.15em,0.4em+.15em) -- (7.8em-.15em,0.4em-.15em);
			\draw[fill=none,  line width=0.05em] (7.8em+.15em,0.4em-.15em) -- (7.8em-.15em,0.4em+.15em);
			\draw[fill=none,  line width=0.05em] (7.2em+.15em,0.4em+.15em) -- (7.2em-.15em,0.4em-.15em);
			\draw[fill=none,  line width=0.05em] (7.2em+.15em,0.4em-.15em) -- (7.2em-.15em,0.4em+.15em);
		}
		\ +  \ \tikz [baseline=0.6em] {
			\draw[fill=none,  line width=0.05em] (7.2em,.40em) circle (0.6em and 0.6em);
			\draw[fill=none,  line width=0.05em] (7.2em,1.6em) circle (0.6em and 0.6em);
			\draw[fill=none, dash pattern=on .1em off .1em, line width=0.05em] (7.2em,1.0em) -- (7.2em,-.2em);
			\draw[fill=black] (7.2em,1.0em) circle (0.10em and 0.10em);
			\draw[fill=black] (7.2em,-0.2em) circle (0.10em and 0.10em);
			\draw[fill=none,  line width=0.05em] (7.2em+.15em,2.2em+.15em) -- (7.2em-.15em,2.2em-.15em);
			\draw[fill=none,  line width=0.05em] (7.2em+.15em,2.2em-.15em) -- (7.2em-.15em,2.2em+.15em);
			\draw[fill=none,  line width=0.05em] (6.6em+.15em,0.4em+.15em) -- (6.6em-.15em,0.4em-.15em);
			\draw[fill=none,  line width=0.05em] (6.6em+.15em,0.4em-.15em) -- (6.6em-.15em,0.4em+.15em);
			\draw[fill=none,  line width=0.05em] (7.8em+.15em,0.4em+.15em) -- (7.8em-.15em,0.4em-.15em);
			\draw[fill=none,  line width=0.05em] (7.8em+.15em,0.4em-.15em) -- (7.8em-.15em,0.4em+.15em);
			\draw[fill=none,  line width=0.05em] (7.2em+.15em,0.4em+.15em) -- (7.2em-.15em,0.4em-.15em);
			\draw[fill=none,  line width=0.05em] (7.2em+.15em,0.4em-.15em) -- (7.2em-.15em,0.4em+.15em);
		} \ + \ \tikz [baseline=0.2em] {
			\draw[fill=none,  line width=0.05em] (7.2em,.40em) circle (0.6em and 0.6em);
			\draw[fill=none,  line width=0.05em] (7.2em,1.6em) circle (0.6em and 0.6em);
			\draw[fill=none,  line width=0.05em] (7.2em,-0.8em) circle (0.6em and 0.6em);
			\draw[fill=none, dash pattern=on .1em off .1em, line width=0.05em] (7.2em,1.0em) -- (7.2em,-.2em);
			\draw[fill=black] (7.2em,1.0em) circle (0.10em and 0.10em);
			\draw[fill=black] (7.2em,-0.2em) circle (0.10em and 0.10em);
			\draw[fill=none,  line width=0.05em] (7.2em+.15em,2.2em+.15em) -- (7.2em-.15em,2.2em-.15em);
			\draw[fill=none,  line width=0.05em] (7.2em+.15em,2.2em-.15em) -- (7.2em-.15em,2.2em+.15em);
			\draw[fill=none,  line width=0.05em] (6.6em+.15em,0.4em+.15em) -- (6.6em-.15em,0.4em-.15em);
			\draw[fill=none,  line width=0.05em] (6.6em+.15em,0.4em-.15em) -- (6.6em-.15em,0.4em+.15em);
			\draw[fill=none,  line width=0.05em] (7.8em+.15em,0.4em+.15em) -- (7.8em-.15em,0.4em-.15em);
			\draw[fill=none,  line width=0.05em] (7.8em+.15em,0.4em-.15em) -- (7.8em-.15em,0.4em+.15em);
			\draw[fill=none,  line width=0.05em] (7.2em+.15em,-1.4em+.15em) -- (7.2em-.15em,-1.4em-.15em);
			\draw[fill=none,  line width=0.05em] (7.2em+.15em,-1.4em-.15em) -- (7.2em-.15em,-1.4em+.15em);
			\draw[fill=none,  line width=0.05em] (7.2em+.15em,0.4em+.15em) -- (7.2em-.15em,0.4em-.15em);
			\draw[fill=none,  line width=0.05em] (7.2em+.15em,0.4em-.15em) -- (7.2em-.15em,0.4em+.15em);
		} \\
		&= -\frac{16\pi^2\bar{\eta}^2\varphi^2 \left(\varphi^2 - \frac{N M}{4\pi} \right)^2 }{ N^3 \varepsilon}  \\
		&\quad - \frac{64\pi^2 \bar{\eta}^2 \varphi^2 \left(\varphi^2 - \frac{N M}{4\pi} \right)^2  }{ N^3}\left[\frac12 - \log \left(\frac{2M + \tilde{M}}{\mu}\right) \right]\\
		&\quad + \frac{16\pi \bar{\eta}^2 M \varphi^2 \left(\varphi^2 - \frac{N M}{4\pi} \right) }{ N^2} \left[1  - \log 2 - \log \frac{M}{\mu}\right]\,,   \\
	\end{aligned}
\end{equation}
while the remaining ones are finite and simply read
\begin{equation}\label{eq:Rtilde1n}
	\tilde{R}_{1,l} = -\frac{8 M^2 \varphi^2}{\pi  N} \left[\bar{\eta} \left(1 - \frac{4\pi \varphi^2}{N M} \right)\right]^{l+1} \int_0^\infty\!\!\!\!\frac{\arctan^l(z) \,\mathrm{d}z }{z^{l-2}\left(z^2 + \frac{\tilde{M}^2}{4 M^2}\right)}
\end{equation}
for $l\geq2$. 
Returning to the superfamily $\tilde{R}_n$, all graphs $n\geq 2$ are UV finite as well. Inserting \eq{dynamical-mass} and \eq{Mtilde}, they can be written as
\begin{equation}\label{eq:Rtilden}
	\begin{aligned}
		\tilde{R}_n &= -\frac{256 \pi\bar{\eta}^{3/2} \varphi^6}{(1 + \sqrt{\bar{\eta}})^3 N^3} \times \\
		& \int_0^\infty\!\!\!\!\mathrm{d}z \frac{z^2}{n}\left[\frac{2\sqrt{\bar{\eta}} (1+\sqrt{\bar{\eta}}) \arctan z}{( 4 z^2  + 5 + 4 \sqrt{\bar{\eta}})(\sqrt{\bar{\eta}} \arctan z + z)}\right]^n .
	\end{aligned}
\end{equation}
Summing the results of \eq{Rn_sol}, \eq{Rtilde1n} and \eq{Rtilden} gives
\begin{equation}\label{eq:FF-sum}
	\sum_{n=4}^\infty R_n + \sum_{n=2}^\infty \left(\tilde{R}_{1,n} + \tilde{R}_{n}\right) = \frac{128 \pi\bar{\eta}^{3/2} \varphi^6}{(1 + \sqrt{\bar{\eta}})^3 N^3} F\left(\sqrt{\bar{\eta}}\right)\,,
\end{equation} 
where $F\left(\sqrt{\bar{\eta}}\right) $ is a function that can be evaluated numerically
\begin{equation}\label{eq:FF}
	\begin{aligned}
		F(x) =  &\int_0^\infty\!\!\!\!\mathrm{d}z  \left[ -\frac{x z (1+ 4 z^2) \arctan z}{5 + 4 x + 4 z^2} + \frac{x^2}{2} \arctan^2 z \right. \\
		&- \frac{x^3}{3 z} \arctan^3 z - z^2 \log \left(1 + \frac{x}{z} \arctan z\right) \\
		&\left.+ 2 z^2 \log \left(1 + \frac{x(3+2x+4z^2)}{z(5+4x+4z^2)} \arctan z \right) \right].
	\end{aligned}
\end{equation}

Finally, all UV divergencies arising at NLO are subtracted minimally by computing LO vacuum graphs with counterterm insertions. The scalar field variable $\varphi$ has a counterterm $\delta \varphi = \mathcal{O}(N^{-2})$, which only contributes to the effective potential at N$^2$LO in large-$N$. On the other hand, the sextic coupling is renormalised via $\bar{\eta} \mapsto \bar{\eta} + \delta \bar{\eta}/\varepsilon$ with
\begin{equation}\label{eq:etabar-ct}
	\delta \bar{\eta} = \frac{6\bar{\eta}^2}{ N }\left(1 - \frac{\pi^2 }{24}\bar{\eta}\right) + \mathcal{O}(N^{-2})\,,
\end{equation}
which includes two- and four-loop contributions~\cite{Pisarski:1982vz}, while higher loops and higher poles are of order $\mathcal{O}(N^{-2})$.
This implies a counterterm from the tree-level potential $\delta\eta \varphi^6 / (6! \varepsilon)$ as well as mass, quartic and sextic couplings to $G_i$, marked as~\tikz [baseline=-0.3em] {
	\draw[fill=white, line width=0.05em] (0em,0em) circle (0.3em and 0.3em);
	\draw[fill=none,  line width=0.05em] (0em+.15em,0em+.15em) -- (0em-.15em,0em-.15em);
	\draw[fill=none,  line width=0.05em] (0em+.15em,0em-.15em) -- (0em-.15em,0em+.15em);
}~below.
The counterterm potential reads
\begin{equation}\label{eq:Vct}
	\begin{aligned}
		V_\text{ct} &= \frac{(4\pi)^2 \varphi^6}{6 } \frac{\delta \bar{\eta}}{\varepsilon N^2}
		+ 
		\tikz [baseline=0em] {
			\draw[fill=none, line width=0.05em] (7.2em,.40em) circle (0.6em and 0.6em);
			\draw[fill=none,  line width=0.05em] (7.2em+.15em,-0.2em+.15em) -- (7.2em-.15em,-0.2em-.15em);
			\draw[fill=none,  line width=0.05em] (7.2em+.15em,-0.2em-.15em) -- (7.2em-.15em,-0.2em+.15em);
			\draw[fill=white, line width=0.05em] (7.2em,1.0em) circle (0.3em and 0.3em);
			\draw[fill=none,  line width=0.05em] (7.2em+.15em,1.0em+.15em) -- (7.2em-.15em,1.0em-.15em);
			\draw[fill=none,  line width=0.05em] (7.2em+.15em,1.0em-.15em) -- (7.2em-.15em,1.0em+.15em);
			} + \tikz [baseline=0em] {
				\draw[fill=none, line width=0.05em] (8.3em,1.0em) circle (0.6em and 0.6em);
				\draw[fill=none, line width=0.05em] (8.3em,-.2em) circle (0.6em and 0.6em);
				\draw[fill=none,  line width=0.05em] (8.3em+.15em,-0.8em+.15em) -- (8.3em-.15em,-0.8em-.15em);
				\draw[fill=none,  line width=0.05em] (8.3em+.15em,-0.8em-.15em) -- (8.3em-.15em,-0.8em+.15em);
				\draw[fill=none,  line width=0.05em] (8.3em+.15em,1.6em+.15em) -- (8.3em-.15em,1.6em-.15em);
				\draw[fill=none,  line width=0.05em] (8.3em+.15em,1.6em-.15em) -- (8.3em-.15em,1.6em+.15em);
				\draw[fill=white] (8.3em,.4em) circle (0.3em and 0.3em);
				\draw[fill=none,  line width=0.05em] (8.3em+.15em,0.4em+.15em) -- (8.3em-.15em,0.4em-.15em);
				\draw[fill=none,  line width=0.05em] (8.3em+.15em,0.4em-.15em) -- (8.3em-.15em,0.4em+.15em);
			} + \tikz [baseline=0em] {
				\draw[fill=none, line width=0.05em] (7.2em,.40em) circle (0.6em and 0.6em);
				\draw[fill=none, line width=0.05em] (8.3em,1.0em) circle (0.6em and 0.6em);
				\draw[fill=none, line width=0.05em] (8.3em,-.2em) circle (0.6em and 0.6em);
				\draw[fill=white] (7.92em,.4em) circle (0.3em and 0.3em);
				\draw[fill=none,  line width=0.05em] (7.92em+.15em,0.4em+.15em) -- (7.92em-.15em,0.4em-.15em);
				\draw[fill=none,  line width=0.05em] (7.92em+.15em,0.4em-.15em) -- (7.92em-.15em,0.4em+.15em);
				\draw[fill=none,  line width=0.05em] (6.6em+.15em,0.4em+.15em) -- (6.6em-.15em,0.4em-.15em);
				\draw[fill=none,  line width=0.05em] (6.6em+.15em,0.4em-.15em) -- (6.6em-.15em,0.4em+.15em);
				\draw[fill=none,  line width=0.05em] (8.3em+.15em,1.6em+.15em) -- (8.3em-.15em,1.6em-.15em);
				\draw[fill=none,  line width=0.05em] (8.3em+.15em,1.6em-.15em) -- (8.3em-.15em,1.6em+.15em);
				\draw[fill=none,  line width=0.05em] (8.3em+.15em,-0.8em+.15em) -- (8.3em-.15em,-0.8em-.15em);
				\draw[fill=none,  line width=0.05em] (8.3em+.15em,-0.8em-.15em) -- (8.3em-.15em,-0.8em+.15em);
			} \\
		&= \frac{(4\pi)^2}{6 }  \left(\varphi^2 - \frac{N M}{4\pi}\right)^3 \frac{\delta \bar{\eta}}{\varepsilon N^2} \\
		&\phantom{=\ }- \frac{ 4\pi \delta\bar{\eta} M }{N}\left(\varphi^2 - \frac{N M}{4\pi}\right)^2\left(1 - \log 2 - \log \frac{M}{\mu}\right)\,.
	\end{aligned}
\end{equation}
The expression \eq{Vct} exactly subtracts the $1/\varepsilon$ poles in the UV divergent ring diagrams \eq{R2}, \eq{R3} and \eq{R1}. The finite term of \eq{Vct} stems from the product of the counterterm and the $\mathcal{O}(\varepsilon)$ contributions of the diagrams. The logarithmic terms $\log \mu$ in \eq{R2}, \eq{R3}, \eq{R1} and \eq{Vct} exactly cancel the RG running of \eq{Veff-LO} as expected.

Overall, the effective potential explicitly depends on the renormalisation scale $\mu$ at NLO, which signals the breaking of scale invariance by quantum fluctuations. As the potential is overall invariant under a renormalisation group transformation, we choose to relate its scale to the field value
\begin{equation}\label{eq:mu-h^2}
	\mu = \frac{4\pi}{N} \varphi^2 \,.
\end{equation}
This choice resums higher order operators, usually contained in logarithmic terms $\log \varphi^2/\mu$ into the coupling $\bar{\eta}$. In consequence, evolving the effective potential to a different field value $\varphi$ involves an RG transformation of $\bar{\eta}(4\pi\varphi^2/N)$.
Due to \eq{eta-bar-beta} being positive until the UV fixed point, studying the potential at higher values of $\bar{\eta}(4\pi \varphi^2/N)$ is surrogate to evaluating $V_\text{eff}/\varphi^6$ at larger field values $\varphi$. 

Thus, we obtain the effective potential at NLO 
\begin{equation}\label{eq:Veff-NLO}
	\begin{aligned}
		V_\text{eff} &= \frac{(4\pi)^2 \varphi^6}{6 N^2} \frac{\bar{\eta}}{(1 + \sqrt{\bar{\eta}})^2}  \ + \ \left.\frac{(4\pi)^2 \varphi^6}{N^3}\frac{\bar{\eta}^{3/2} }{(1 + \sqrt{\bar{\eta}})^3} \right[ \\
		&\qquad + \frac13 - \sqrt{\bar{\eta}} - \frac13\left(5 + 4 \sqrt{\bar{\eta}}\right)^{3/2} - \frac72 \zeta_3 \bar{\eta}^{3/2}\\
		&\qquad + 4 \bar{\eta}\left(1 - 2\log2\right) + \frac{\pi^2}{12}\bar{\eta}^{3/2}\left(1 - \log2\right)\\
		&\qquad- 4 \sqrt{\bar{\eta}} \left(1 - \frac{\pi^2}{12}\bar{\eta}\right) \log \left(1 + \frac1{\sqrt{\bar{\eta}}}\right) \\
		&\qquad+  4\sqrt{\bar{\eta}}\left(1 + \sqrt{\bar{\eta}}\right)\log \left(2 + \sqrt{5 + 4 \sqrt{\bar{\eta}}}\right) 
		\\
	&\qquad \left.+ \frac{8}{\pi} F\left(\sqrt{\bar{\eta}}\right) \right] + \mathcal{O}(N^{-4}) \,,
	\end{aligned}
\end{equation} 
where $F\left(\sqrt{\bar{\eta}}\right) $ is defined in \eq{FF}.
\begin{figure}
	\centering
	\includegraphics[width=\columnwidth]{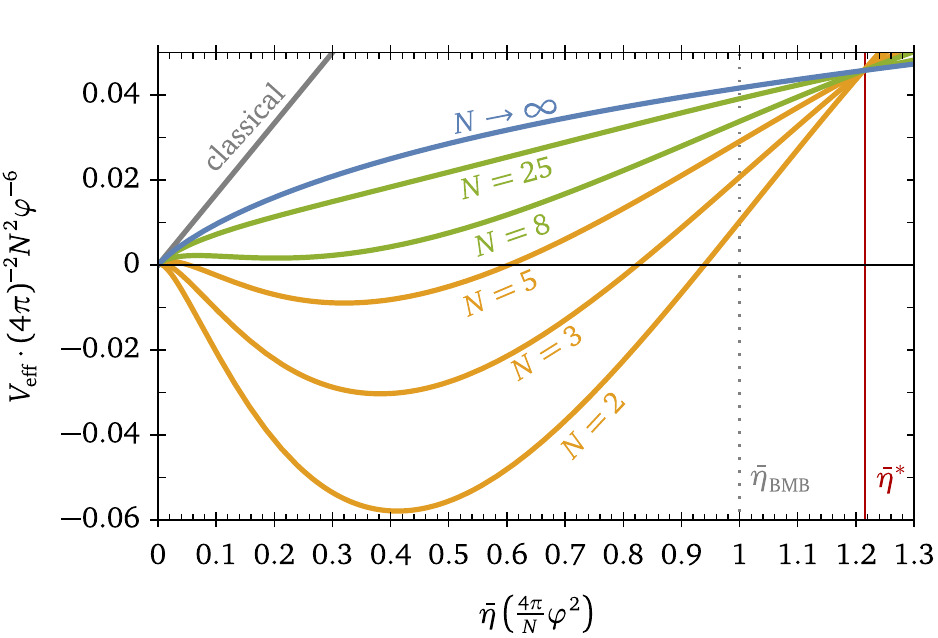}
	\caption{
		Sextic term of the classical (solid grey) and next-to-leading order effective potential~\eq{Veff-NLO} as a function of the rescaled sextic coupling as a function of $\bar{\eta}$, evaluated at the renormalisation scale $\mu = 4\pi \varphi^2/N$, for various choices of $N$.  
		The limit $N \to \infty$ (blue) singles out the LO effective potential~\eq{Veff-LO}, which has a smaller but stable sextic coupling compared to the classical case (solid grey). With decreasing $N$, the potential remains stable (green) until it turns unstable at integer values $N < 8$ (yellow). However, the large-$N$ expansion might not be reliable in this case.
		The BMB endpoint $\bar{\eta}_\text{BMB} = 1$ (grey dashed) and Pisarski's UV fixed point~\eq{Pisarski-bar} $\bar{\eta}^* \approx 12/\pi^2$ (red) are also shown.
	}
	\label{fig:Veff-etabar}
\end{figure}
Eq.~\eq{Veff-NLO} is one of the main results of this work and also displayed in \fig{Veff-etabar}. 
For integer values $N \geq 8$, the potential is stable in the sense that it is both bounded from below and that $\varphi = 0$ is an absolute minimum.
At $N < 8$, the potential is still bounded from below but develops a deeper minimum $\varphi_\text{min} \neq 0$, hinting at spontaneous symmetry breaking. 
However, as \eq{Veff-NLO} has been obtained in a large-$N$ expansion, the prediction might not be reliable in this case. 

The effective potential can be extended na\"ively to Pisarski's UV fixed point $\bar{\eta}^* \approx 12/\pi^2$~\eq{Pisarski-bar}, marked red in \fig{Veff-etabar}. All curves appear to be converging at this point as the NLO corrections are numerically small. 
While the UV fixed point is stable for sufficient large $N$, it has been argued that an instability occurs through the non-perturbative BMB phenomenon already at $\bar{\eta}_\text{BMB}=1$~\cite{Bardeen:1983rv}, shown as grey dotted line in \fig{Veff-etabar}. We will turn towards this effect in the next section.

\section{Bardeen-Moshe-Bander Phenomenon}\label{sec:BMB}

The BMB phenomenon~\cite{Bardeen:1983rv} is a non-perturbative effect that breaks the scale invariance spontaneously through dimensional transmutation at sextic coupling values
\begin{equation}\label{eq:eta_BMB}
	\bar{\eta}_\text{BMB} = 1\,.
\end{equation}
This terminates the line of accidental scale invariance at leading order large-$N$. Moreover, the proper UV fixed point~\eq{Pisarski-bar} lies in the affected parameter region $\bar{\eta}^* > \bar{\eta}_\text{BMB}$. 
The spontaneous breaking of scale invariance at the BMB point gives rise to a massless Nambu-Goldstone boson, the \textit{dilaton}~\cite{Bardeen:1983rv}.
In the following, we review the BMB phenomenon at LO and extend its analysis to NLO in the $1/N$-expansion.

\subsection{Leading Order}

Capturing the onset of the BMB is challenging in perturbation theory and especially dimensional regularisation. Nevertheless, the generation of a dynamical mass $\mathscr{M}$ by quantum effects of the initially massless field $\phi$ can be understood from a gap equation 
\begin{equation}\label{eq:gap-BMB}
	\begin{tikzpicture}[baseline=-.3em]
			\draw[fill=none, line width=0.05em] (-1.0em,0) -- (0em,0);
			\draw[fill=none, line width=0.05em] (1.0em,0) -- (0em,0);
			\draw[fill=none, line width=0.05em] (-.15em,-0.15em) -- (+.15em,0.15em);
			\draw[fill=none, line width=0.05em] (+.15em,-0.15em) -- (-.15em,0.15em);
		\end{tikzpicture} \quad= \quad \begin{tikzpicture}[baseline=-.3em]
			\draw[fill=none, line width=0.05em] (-1.0em,0) -- (0em,0);
			\draw[fill=none, line width=0.05em] (1.0em,0) -- (0em,0);
		\end{tikzpicture}  \quad +\quad \begin{tikzpicture}[baseline=-.3em]
			\draw[fill=none, line width=0.05em] (-1.5em,0) -- (0,0);
			\draw[fill=none, line width=0.05em] (+1.5em,0) -- (0,0);
			\draw[fill=none, line width=0.05em] (0em,-.6em) circle (0.6em and 0.6em);
			\draw[fill=none, line width=0.05em] (0em,+.6em) circle (0.6em and 0.6em);
			\draw[fill=none, line width=0.05em] (-.15em,1.35em) -- (+.15em,1.05em);
			\draw[fill=none, line width=0.05em] (+.15em,1.35em) -- (-.15em,1.05em);
			\draw[fill=none, line width=0.05em] (-.15em,-1.35em) -- (+.15em,-1.05em);
			\draw[fill=none, line width=0.05em] (+.15em,-1.35em) -- (-.15em,-1.05em);
			\draw[fill=black] (0,0) circle (0.15em and 0.15em);
			\draw[fill=none, line width=0.05em] (+.8em,-0.15em) -- (+1.1em,+0.15em);
			\draw[fill=none, line width=0.05em] (+.8em,+0.15em) -- (+1.1em,-0.15em);
		\end{tikzpicture}
	\end{equation}
	which encodes the resummation of all leading large-$N$ contributions to the two-point function $\braket{T \phi_i(x) \phi_j(y)}$. Here, the uncrossed lines denote the massless propagators of $\phi$, while the crossed ones represent dressed ones with a putative mass $\mathscr{M}$. This gap equation yields a self-consistency condition
	\begin{equation}\label{eq:gap-BMB-condition}
		\mathscr{M}^2 = \bar{\eta} \mathscr{M}^2\,,
	\end{equation}
	which for $\bar{\eta} < \bar{\eta}_\text{BMB}$ has only the solution $\mathscr{M} = 0$. However, for $\bar{\eta} = \bar{\eta}_\text{BMB}$, the condition is also valid for any $\mathscr{M} \neq 0$, and a dynamical mass is generated at loop level.
	This leads to the formation a scalar condensate as evident from the two-point correlator
	\begin{equation}\label{eq:correlator}
		\braket{T \phi_k(x)\phi_k(x)} = \int \frac{\mathrm{d}^3k}{(2\pi)^3} \frac{i N}{k^2 - \mathscr{M}^2} = - \frac{N |\mathscr{M}|}{4\pi}\,,
	\end{equation} 
	while the global symmetry remains unbroken as $\braket{\phi} = 0$. Accounting for the formation of a condensate by shifting $\phi^2$ around $\braket{\phi^2}$ yields a \textit{trivial} gap equation
	\begin{equation}
		\mathscr{M}^2 =  \bar{\eta}\mathscr{M}^2 + \frac{\eta}{120} \braket{\phi^2}^2 - \frac{N \eta}{240\pi} \mathscr{M} \braket{\phi^2} + \frac{N^2 \eta}{1920\pi^2} \mathscr{M}^2
	\end{equation}
	as the last three terms cancel against each other.
	
	The actual value of the condensate $\braket{\phi^2}$ or equivalent the dynamical mass scale $\mathscr{M}$ are determined by minimising the effective potential.
	To this end, we return to the 2PI potential $V_\text{2PI}(\varphi,M)$~\eq{V_2PI}. In \Sec{Veff}, we used this formalism as a vehicle to obtain the effective potenital $V_\text{eff}(\varphi) = V_\text{2PI}(\varphi,\,M_+(\varphi))$ by inserting the minimum $M_+(\varphi)$, see \eq{dynamical-mass}.
	Here, we retain $V_\text{2PI}(\varphi,M)$ thus treating the field $\varphi$ and the Goldstone dynamical mass $M$ as independent quantities. This allows us to identify a different set of vacua. 

\begin{figure}
		\centering
		\includegraphics[scale=.4]{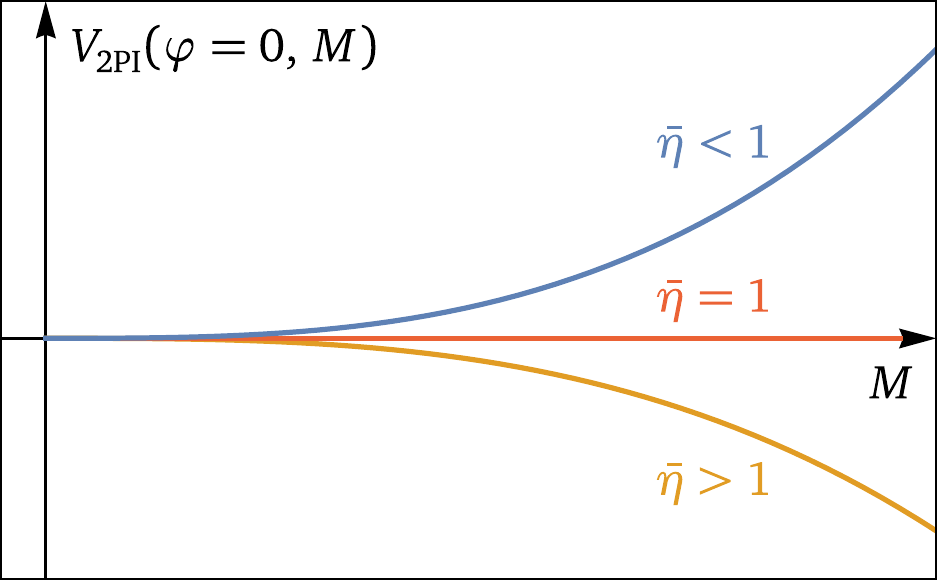}
		\caption{Leading-order 2PI potential~\eq{BMB-stability} at $\varphi=0$ and as a function of $M$. For $\bar{\eta} < 1$, the potential is bounded from below and has a minimum at $M=0$, while it is unstable for $\bar{\eta} > 1$. At $\bar{\eta} = 1$, $V_\text{2PI}(0,M)$ is flat and the BMB phenomenon occurs. }
		\label{fig:V2PI-M-0}
	\end{figure}

	The potential $V_\text{2PI}$~\eq{V_2PI} is shown in \fig{V2PI-M} with various values of $\bar{\eta}$ and fixed $\varphi$.
	For $\bar{\eta} < 1$, there is a minimum $\varphi = M = 0$ which is bounded from below and hence a stable ground state.
	At $\bar{\eta} = 1$, the minimum turns into a saddle point as the potential becomes flat in the direction of $M$. This is most obvious when projecting $V_\text{2PI}$ onto the section $\varphi = 0$, which yields the expression
	\begin{equation}\label{eq:BMB-stability}
		V_\text{2PI}(0,M) = \frac{N M^3 }{24\pi}(1 - \bar{\eta})\,,
	\end{equation}
	that was also obtained in \cite{Gudmundsdottir:1984rr} and is displayed in \fig{V2PI-M-0}. 
	At $\bar{\eta} = 1$, \eq{BMB-stability} is flat and any value of $M = \mathscr{M}$ is a valid ground state into which the system can transition, dynamically generating a scale $\mathscr{M}$ and thus breaking scale invariance. As the scale is arbitrary, we were unable to determine a fixed value in \eq{gap-BMB}. 
	The composite field 
	\begin{equation}\label{eq:Dilaton}
		D(x) = \frac{1}{N}\phi_i(x) \phi_i(x)
	\end{equation}
	acts as a dilaton mode that shifts the ground state to different values of $\mathscr{M}$, and is massless at the BMB due to the flatness of the 2PI potential~\eq{BMB-stability}. 
	The masslessness can be seen explicitly from the LO dilaton propagator~\cite{Bardeen:1983rv} in the $O(N)$ symmetric theory 
	\begin{equation}
		\begin{aligned}
		\braket{0|D(p)D(-p)|0} &= \sum_{n=0}^\infty 
		\tikz [baseline=-0.1em]{
			\draw[fill=black] (+0.8em,0em) circle (0.15em and 0.15em);
			\draw[fill=black] (-0.8em,0em) circle (0.15em and 0.15em);
			\draw[fill=none, line width=0.05em] (0em,0em) circle (0.70em and 0.40em);
			\draw[fill=none, line width=0.05em] (1.6em,0.4em) arc [start angle=90, end angle=270, x radius=0.70em, y radius=0.4em];
			\draw[fill=none, line width=0.05em] (-1.6em,0.4em) arc [start angle=90, end angle=-90, x radius=0.70em, y radius=0.4em];
			\draw[fill=none, line width=0.05em] (-0.8em,0.75em) circle (0.3em and 0.6em);
			\draw[fill=none, line width=0.05em] (0.8em,0.75em) circle (0.3em and 0.6em);

			\draw[fill=none, line width=0.05em] (1.5em,0.4em) -- (1.8em,0.4em);
			\draw[fill=none, line width=0.05em] (1.5em,-0.4em) -- (1.8em,-0.4em);
			\draw[fill=none, line width=0.05em] (-1.5em,0.4em) -- (-1.8em,0.4em);
			\draw[fill=none, line width=0.05em] (-1.5em,-0.4em) -- (-1.8em,-0.4em);
			
			\draw[fill=none,  line width=0.05em] (0em+.15em,.4em+.15em) -- (0em-.15em,.4em-.15em);
			\draw[fill=none,  line width=0.05em] (0em+.15em,.4em-.15em) -- (0em-.15em,.4em+.15em);	
			\draw[fill=none,  line width=0.05em] (0em+.15em,-.4em+.15em) -- (0em-.15em,-.4em-.15em);
			\draw[fill=none,  line width=0.05em] (0em+.15em,-.4em-.15em) -- (0em-.15em,-.4em+.15em);

			\draw[fill=none,  line width=0.05em] (0.5em+.15em,0.7em+.15em) -- (0.5em-.15em,0.7em-.15em);
			\draw[fill=none,  line width=0.05em] (0.5em+.15em,0.7em-.15em) -- (0.5em-.15em,0.7em+.15em);	
			\draw[fill=none,  line width=0.05em] (-1.1em+.15em,0.7em+.15em) -- (-1.1em-.15em,0.7em-.15em);
			\draw[fill=none,  line width=0.05em] (-1.1em+.15em,0.7em-.15em) -- (-1.1em-.15em,0.7em+.15em);	
			\node[anchor=east] at (+0.9em,0.3em) {$\Bigg[^{\phantom{n}}$};
			\node[anchor=west] at (+0.5em,0.3em) {$\Bigg]^n$};
		} \\
		&= \frac{16 \pi \bar{\eta} \mathscr{M}}{N \left[1 - \bar{\eta}\frac{2\mathscr{M}}{p} \arctan \frac{p}{2\mathscr{M}}\right]} \\
%		&= \frac{16 \pi \bar{\eta} \mathscr{M}}{N \left[1 - \bar{\eta} + \frac{\bar{\eta} p^2}{12 \mathscr{M}^2} \right]} + \mathcal{O}(p^4)\\
		&= \frac{192 \pi}{N} \frac{\mathscr{M}^3}{p^2 + m_D^2 } + \mathcal{O}(p^4)
	\end{aligned}
	\end{equation}
	with Euclidean momentum $p$. We read off the dilaton mass 
	\begin{equation}\label{eq:Dilaton-mass}
		m_D^2 = 12\left(\frac{1- \bar{\eta}}{\bar{\eta}}\right)\mathscr{M}^2 \,,
	\end{equation}
	which indeed vanishes at $\bar{\eta} = 1$.
	The effective potential at the BMB point, i.e.
	\begin{equation}
		V_\text{eff} = \frac{1}{2} \mathscr{M}^2 \varphi^2 - \frac{2\pi}{N} \mathscr{M} \varphi^4 + \frac{8\pi^2}{3 N^2} \varphi^6\
	\end{equation}
	has a stable ground state at $\varphi=0$. Thus, the vacua for each value $M=\mathscr{M}$ are all degenerate. Moreover, the potential is bounded from below. This is in agreement with the FRG results~\cite{Litim:2017cnl}, though the exact shape of the potential differs. This is to be expected as this detail appears to be scheme dependent already within the FRG~\cite{Litim:2017cnl}.  
	 
	At $\bar{\eta} > 1$, the former ground state $M=\varphi=0$ becomes a saddle point unstable in the $M$-direction, and the potential is not bounded from below. 
	In the next section, we will find that this is not a sign of sickness, but is merely an artefact of the LO approximation. 

	We close our LO discussion by pointing out that each point of $\bar{\eta}$ corresponds to a separate theory since there is no RG evolution. Thus, there is no transition from a stable $(\bar{\eta} < 1)$ to an unstable regime $(\bar{\eta} \geq 1)$. In particular, while the true UV fixed point~\eq{Pisarski-bar} lies in the unstable regime, one cannot determine whether the BMB phenomenon impedes the RG evolution towards this fixed point at LO in large-$N$. To do so, next-to-leading accuracy is required. 

	\subsection{Next-to-Leading Order}
	Now, we investigate the BMB phenomenon at NLO in the large-$N$ expansion by computing corrections to $V_\text{2PI}$~\eq{BMB-stability}.
	For simplicity, we work at the field minimum $\varphi=0$ where the BMB instability occurs. This restores the full $O(N)$ symmetry, and we can treat all components of $\phi_k$ on equal footing. 
	In general, the 2PI potential $V_\text{2PI}$ is a functional of the dressed propagator $\chi_{kl}$ with $k,l=1,\dots,\,N$. Both of them can be expanded in the large-$N$ limit
	\begin{equation}\label{eq:V2PI-NLO}
		V_\text{2PI} = V_\text{2PI}^\text{LO}[\chi^\text{LO} ] +  V_\text{2PI}^\text{NLO}[\chi^\text{LO} ] + \mathrm{Tr}\left[ \frac{\delta V_\text{2PI}^\text{LO}[\chi^\text{LO} ]}{\delta \chi^\text{LO}} \chi^\text{NLO}\!\!\right] ,
	\end{equation}
	where $\mathrm{Tr}[...]$ is understood as a combined summation of indices and integration of momenta. 
	Starting from $V_\text{2PI}$, the effective potential is obtained by inserting the stationary value of $\chi$
	\begin{equation}
		0 = \frac{\delta V_\text{2PI}[\chi]}{ \delta \chi} = \frac{\delta V_\text{2PI}^\text{LO}[\chi^\text{LO} ]}{\delta \chi^\text{LO}} + \mathcal{O}(\text{NLO}) \,,
	\end{equation}
	which suggests that the third term in \eq{V2PI-NLO} is actually of order N$^2$LO. Inserting the LO propagator
	\begin{equation}
		\chi^\text{LO}_{kl}(p) = \frac{i \delta_{kl}}{p^2 - \mathscr{M}^2}
	\end{equation}
	into the 2PI potential yields an effective potential $V_\mathscr{M}$ for the mass parameter $\mathscr{M}$.
	Explicitly, we obtain 
	\begin{equation}
		V_\mathscr{M} = \frac{N \mathscr{M}^3}{24\pi} + \tikz [baseline=0em] {
			\draw[fill=none, line width=0.05em] (7.2em,.40em) circle (0.6em and 0.6em);
			\draw[fill=none, line width=0.05em] (8.3em,1.0em) circle (0.6em and 0.6em);
			\draw[fill=none, line width=0.05em] (8.3em,-.2em) circle (0.6em and 0.6em);
			\draw[fill=black] (7.92em,.4em) circle (0.15em and 0.15em);
			\draw[fill=none,  line width=0.05em] (6.6em+.15em,0.4em+.15em) -- (6.6em-.15em,0.4em-.15em);
			\draw[fill=none,  line width=0.05em] (6.6em+.15em,0.4em-.15em) -- (6.6em-.15em,0.4em+.15em);
			\draw[fill=none,  line width=0.05em] (8.3em+.15em,1.6em+.15em) -- (8.3em-.15em,1.6em-.15em);
			\draw[fill=none,  line width=0.05em] (8.3em+.15em,1.6em-.15em) -- (8.3em-.15em,1.6em+.15em);
			\draw[fill=none,  line width=0.05em] (8.3em+.15em,-0.8em+.15em) -- (8.3em-.15em,-0.8em-.15em);
			\draw[fill=none,  line width=0.05em] (8.3em+.15em,-0.8em-.15em) -- (8.3em-.15em,-0.8em+.15em);
		} +  \tikz [baseline=0em] {
			\draw[fill=none, line width=0.05em] (7.2em,.40em) circle (0.6em and 0.6em);
			\draw[fill=none, line width=0.05em] (8.3em,1.0em) circle (0.6em and 0.6em);
			\draw[fill=none, line width=0.05em] (8.3em,-.2em) circle (0.6em and 0.6em);
			\draw[fill=white] (7.92em,.4em) circle (0.3em and 0.3em);
			\draw[fill=none,  line width=0.05em] (7.92em+.15em,0.4em+.15em) -- (7.92em-.15em,0.4em-.15em);
			\draw[fill=none,  line width=0.05em] (7.92em+.15em,0.4em-.15em) -- (7.92em-.15em,0.4em+.15em);
			\draw[fill=none,  line width=0.05em] (6.6em+.15em,0.4em+.15em) -- (6.6em-.15em,0.4em-.15em);
			\draw[fill=none,  line width=0.05em] (6.6em+.15em,0.4em-.15em) -- (6.6em-.15em,0.4em+.15em);
			\draw[fill=none,  line width=0.05em] (8.3em+.15em,1.6em+.15em) -- (8.3em-.15em,1.6em-.15em);
			\draw[fill=none,  line width=0.05em] (8.3em+.15em,1.6em-.15em) -- (8.3em-.15em,1.6em+.15em);
			\draw[fill=none,  line width=0.05em] (8.3em+.15em,-0.8em+.15em) -- (8.3em-.15em,-0.8em-.15em);
			\draw[fill=none,  line width=0.05em] (8.3em+.15em,-0.8em-.15em) -- (8.3em-.15em,-0.8em+.15em);
		} + \sum_{n=2}^\infty R_n \bigg|_{\substack{\!\!\!\varphi=0\\M=\mathscr{M}}}\,,
	\end{equation}
	where~\tikz [baseline=-0.3em] {
		\draw[fill=white, line width=0.05em] (0em,0em) circle (0.3em and 0.3em);
		\draw[fill=none,  line width=0.05em] (0em+.15em,0em+.15em) -- (0em-.15em,0em-.15em);
		\draw[fill=none,  line width=0.05em] (0em+.15em,0em-.15em) -- (0em-.15em,0em+.15em);
	} marks the insertion of a counterterm~\eq{etabar-ct} and
	the last terms have been computed in \eq{R2}, \eq{R3} and \eq{Rn_sol}. Putting all pieces together, we arrive at the potential 
	\begin{equation}\label{eq:VM-M}
		\begin{aligned}
			&V_\mathscr{M} = \frac{N\mathscr{M}^3 (1 - \bar{\eta})}{24\pi} + \frac{\mathscr{M}^3 }{24\pi} \bigg[  - 6 \bar{\eta}\\
			&\qquad  + 12\bar{\eta}^2\left(3 - 4\log2 - 2 \log \frac{\mathscr{M}}{\mu}\right) + 21 \zeta_3 \bar{\eta}^3 \\
			&\qquad - \frac{\pi^2\bar{\eta}^3}2 \left(1 - 2 \log2 - 4\log\frac{\mathscr{M}}{\mu}\right)  + \frac{48}{\pi} f(\bar{\eta})\bigg] \,,
		\end{aligned}
	\end{equation}
	where $f(x)$ is given in \eq{ff1}. For the sake of discussion, we write it in the compact shape
	\begin{equation}\label{eq:V-c12}
		V_\mathscr{M} = \mathscr{M}^3\left[c_0(\bar{\eta}) + c_1(\bar{\eta})\,\log \frac{\mathscr{M}}{\mu}\right]\,.
	\end{equation}
	At LO, the logarithmic term is absent, i.e. $c_1=0$, such that the BMB phenomenon occurs when $c_0(\bar{\eta}_\text{BMB}) = 0$, where the minimum at $\mathscr{M}=0$ is lifted. This yields $\bar{\eta}_\text{BMB} = 1$, and renders $V_\mathscr{M}$ exactly flat at the BMB point. 
	At NLO, $c_0$ obtains corrections, but still has a zero $c_0(\bar{\eta}_0) = 0$ at
	\begin{equation}\label{eq:eta_BMB-NLO}
		\begin{aligned}
			\bar{\eta}_0 &= 1 + \frac{\pi^2}{2N}\left(2 \log2 - 1\right) + \frac{48}{\pi N} f(1) \\
			&\phantom{ = 1 \ }+ \frac{3\left(10 + 7 \zeta_3 - 16 \log 2\right)}{N} + \mathcal{O}\left(N^{-2}\right) \\
			&\approx 1 + \frac{9.0584}{N} + \mathcal{O}\left(N^{-2}\right)\,.
		\end{aligned}
	\end{equation}
	\begin{figure}
		\centering
		\includegraphics[scale=.50]{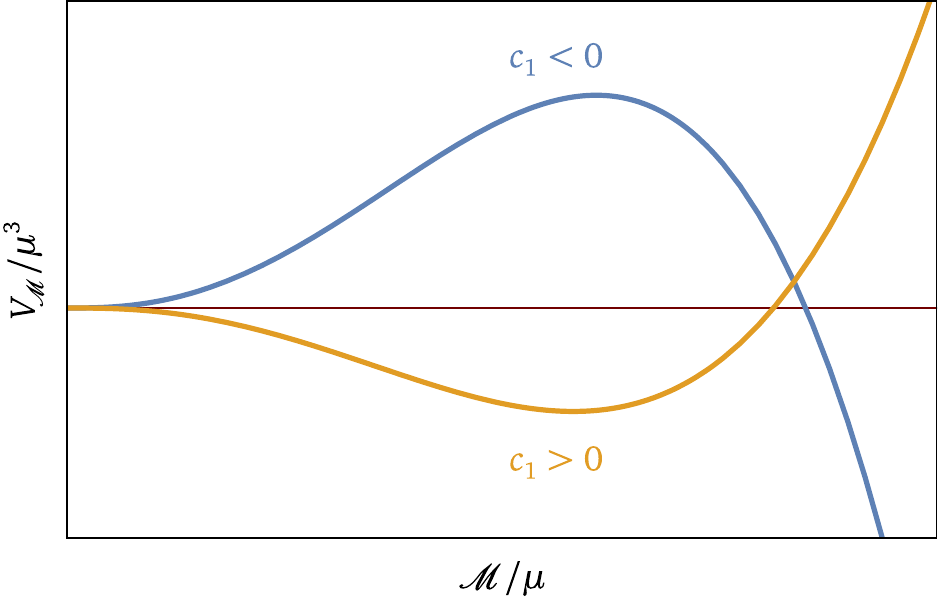}
		\caption{Schematic dependence of the effective potential at next-to leading order in large-$N$~\eq{V-c12} on the sign of $c_1$. For $c_1 > 0$, there is a global minimum at $\mathscr{M} \neq 0$, and $V_\mathscr{M}$ is bounded from below. With $c_1 < 0$, there is a local minimum $\mathscr{M} = 0$, the global maximum lies at $\mathscr{M} > 0$, and $V_\mathscr{M} $ is not bounded from below for $\mathscr{M}/\mu \to \infty$.
		}
		\label{fig:VM-schema}
	\end{figure}
	However, $\bar{\eta}_0$ is not simply the generalisation of $\bar{\eta}_\text{BMB}$ at NLO.
	The stability of \eq{V-c12} around the ground state $\mathscr{M} = 0$ does not hinge on $c_0(\bar{\eta})$ but rather on the sign of $c_1(\bar{\eta})$, as depicted schematically in \fig{VM-schema}. Concretely, because $\log \mathscr{M}/\mu \to - \infty$ as $\mathscr{M} \to 0$, the ground state is stable for $c_1 < 0$. 
	As evident from \eq{VM-M}, the coefficient is related to the $\beta$-function 
	\begin{equation}
		c_1(\bar{\eta}) = - \beta_{\bar{\eta}} < 0
	\end{equation}
	for all $\bar{\eta} < \bar{\eta}^*$, cf.~\eq{eta-bar-beta}. This suggests that $\mathscr{M} = 0$ remains a minimum of $V_\mathscr{M}$ and there is no breaking of scale invariance for $0 \leq \bar{\eta} \leq \bar{\eta}^*$, i.e. the UV fixed point~\eq{Pisarski-bar} exists.
	The appearance of $\beta_{\bar{\eta}}$ in \eq{VM-M}  is rooted in the overall RG invariance of the potential, i.e.
	 \begin{equation}
		0 = \left(\frac{\partial}{\partial \log \mu} + \gamma_{\!\!\mathscr{M}} \frac{\partial}{\partial \log \mathscr{M}}  + \beta_{\bar{\eta}} \frac{\partial}{\partial \bar{\eta}}\right)V_\mathscr{M}\,.
	 \end{equation}
	 Thus, its logarithms may be resummed via 
	 \begin{equation}
		V_\mathscr{M} = \mathscr{M}^3 \exp\left\{\frac{\log\left(\frac{\mathscr{M}}{\mu}\right)}{1-\gamma_{\!\!\mathscr{M}}} \left[3\gamma_{\!\!\mathscr{M}} + \beta_{\bar{\eta}} \frac{\partial}{\partial \bar{\eta}}\right]\right\} c_0(\bar{\eta}) \,,
	 \end{equation}
	 where $\gamma_{\!\!\mathscr{M}} = \mathcal{O}(N^{-2}) $~\cite{Pisarski:1982vz} is the anomalous dimension of the mass operator $\mathscr{M}$ and only contributes at N$^2$LO in the large-$N$ expansion. As a consequence, the logarithmic term arises solely from the LO expression $c_0 = N(1 - \bar{\eta})/(24\pi)$.

	To argue that the theory remains well-defined for $\bar{\eta} \leq 1$, we scrutinise the radius of convergence of 
	the effective potential $V_\mathscr{M}$. 
	A complication arises due to the series of ring diagrams~\eq{Rn_sol}, which are summed into logarithmic series contained in the function $f(\bar{\eta})$ as defined in~\eq{ff1}.
	The radius of convergence for the sum of diagrams is $\bar{\eta} \leq 1$,
	which, strictly speaking, does not permit the exchange the summation and integration as was done to obtain~\eq{ff1}.
	Doing so regardless, we find that $f(\bar{\eta})$ does not exhibit a pole but remains finite as $\bar{\eta} \to 1$. 
	Instead, $f(\bar{\eta})$ develops an imaginary part for $\bar{\eta} > 1$ and so does $V_\mathscr{M}$. 
	Imaginary parts of effective potentials are known to be related to their non-convexity and hint at the coexistance of several ground states. To be precise, the imaginary part is interpreted as a decay rate between well-defined, localised ground states~\cite{Weinberg:1987vp,Sher:1988mj}.	
	The emergence of an imaginary part in our case can be understood from the fact that we have identified two vacua corresponding to the dynamical masses $M_\pm$~\eq{dynamical-mass}, in agreement with~\cite{Matsubara:1987iz}.
	For $\bar{\eta} \leq 1$, only $M_+$ is valid and there is no imaginary part.
	$M_-$ becomes a viable vacuum for $\bar{\eta} > 1$, though it is a maximum of the $V_\text{2PI}$ potential. Thus, $M_\pm$ coexist but $M_+$ corresponds to field configurations with lower energy. Thus, there is no imaginary part for $V_\text{eff}$ computed with $\varphi \neq 0$ in \eq{Veff-NLO}, as expected by~\cite{Townsend:1976sz}. 
	On the other hand, $V_\mathscr{M}$ is determined at $\varphi = 0$ where $M_{\pm}$ become degenerate, thus introducing an imaginary part when both vacua coexist at $\bar{\eta} > 1$. Note that this imaginary part of $V_\mathscr{M}$  vanishes at the ground state, $\mathscr{M} = 0$, which does not suggest an instability.
	Furthermore, the real part of \eq{VM-M} is a smooth function for the coupling values $0 \leq \bar{\eta} \leq \bar{\eta}^*$.
	Thus, the imaginary part does not pose a problem to our conclusion.

	The consequence of our finding is profound: scale invariance cannot be broken by the BMB phenomenon, as it is inherently broken by quantum fluctuations, concretely by the logarithmic term in~\eq{V-c12}.
	Quantum corrections fix the ground state to $\mathscr{M} = 0$, and the effective potential of the mass parameter $\mathscr{M}$ cannot be rendered flat for any value of $\bar{\eta}$. Thus, no dimensional transmutation occurs and the BMB mechanism utterly disappears in the face of NLO corrections.

	A similar conclusion can be drawn by considering the gap equation. At NLO, it reads diagrammatically 
	\begin{equation}\label{eq:gap-BMB-NLO}
		\begin{tikzpicture}[baseline=-.3em]
				\draw[fill=none, line width=0.05em] (-1.0em,0) -- (0em,0);
				\draw[fill=none, line width=0.05em] (1.0em,0) -- (0em,0);
				\draw[fill=none, line width=0.05em] (-.15em,-0.15em) -- (+.15em,0.15em);
				\draw[fill=none, line width=0.05em] (+.15em,-0.15em) -- (-.15em,0.15em);
			\end{tikzpicture}  =   \begin{tikzpicture}[baseline=-.3em]
				\draw[fill=none, line width=0.05em] (-1.em,0) -- (0em,0);
				\draw[fill=none, line width=0.05em] (+1.em,0) -- (0em,0);
			\end{tikzpicture}   +  \begin{tikzpicture}[baseline=-.3em]
				\draw[fill=none, line width=0.05em] (-1.0em,0) -- (0,0);
				\draw[fill=none, line width=0.05em] (+1.0em,0) -- (0,0);
				\draw[fill=none, line width=0.05em] (0em,-.6em) circle (0.6em and 0.6em);
				\draw[fill=none, line width=0.05em] (0em,+.6em) circle (0.6em and 0.6em);
				\draw[fill=none, line width=0.05em] (-.15em,1.35em) -- (+.15em,1.05em);
				\draw[fill=none, line width=0.05em] (+.15em,1.35em) -- (-.15em,1.05em);
				\draw[fill=none, line width=0.05em] (-.15em,-1.35em) -- (+.15em,-1.05em);
				\draw[fill=none, line width=0.05em] (+.15em,-1.35em) -- (-.15em,-1.05em);
				\draw[fill=black] (0,0) circle (0.15em and 0.15em);
				\draw[fill=none, line width=0.05em] (+.6em,-0.15em) -- (+0.9em,+0.15em);
				\draw[fill=none, line width=0.05em] (+.6em,+0.15em) -- (+0.9em,-0.15em);
			\end{tikzpicture} +  \begin{tikzpicture}[baseline=-.3em]
				\draw[fill=none, line width=0.05em] (-1.0em,0) -- (0,0);
				\draw[fill=none, line width=0.05em] (+1.0em,0) -- (0,0);
				\draw[fill=none, line width=0.05em] (0em,-.6em) circle (0.6em and 0.6em);
				\draw[fill=none, line width=0.05em] (0em,+.6em) circle (0.6em and 0.6em);
				\draw[fill=none, line width=0.05em] (-.15em,1.35em) -- (+.15em,1.05em);
				\draw[fill=none, line width=0.05em] (+.15em,1.35em) -- (-.15em,1.05em);
				\draw[fill=none, line width=0.05em] (-.15em,-1.35em) -- (+.15em,-1.05em);
				\draw[fill=none, line width=0.05em] (+.15em,-1.35em) -- (-.15em,-1.05em);
				\draw[fill=white] (0.em,.em) circle (0.3em and 0.3em);
				\draw[fill=none,  line width=0.05em] (0.em+.15em,0.em+.15em) -- (0.em-.15em,0.em-.15em);
				\draw[fill=none,  line width=0.05em] (0.em+.15em,0.em-.15em) -- (0.em-.15em,0.em+.15em);
				\draw[fill=none, line width=0.05em] (+.6em,-0.15em) -- (+0.9em,+0.15em);
				\draw[fill=none, line width=0.05em] (+.6em,+0.15em) -- (+0.9em,-0.15em);
			\end{tikzpicture}
			+  \begin{tikzpicture}[baseline=1.7em,scale=1.3]
				\draw[fill=none, line width=0.05em] (-1.0em,0) -- (0,0);
				\draw[fill=none, line width=0.05em] (+1.0em,0) -- (0,0);
				\draw[fill=black] (0,0) circle (0.10em and 0.10em);
				\draw[fill=none, line width=0.05em, rotate around={50:(0em,0em)}] (0em,0.5em) circle (0.2em and 0.5em);
				\draw[fill=none, line width=0.05em, rotate around={-50:(0em,0em)}] (0em,0.5em) circle (0.2em and 0.5em);
				\draw[fill=black] ({sin(50)*2*.5em},{cos(50)*2*.5em}) circle (0.10em and 0.10em);
				\draw[fill=black] ({-sin(50)*2*.5em},{cos(50)*2*.5em}) circle (0.10em and 0.10em);
				\draw[fill=none, line width=0.05em] ({.5em + sin(50)*2*.5em},{cos(50)*2*.5em}) circle (0.5em and 0.2em);
				\draw[fill=none, line width=0.05em] ({-.5em - sin(50)*2*.5em},{cos(50)*2*.5em}) circle (0.5em and 0.2em);

				\draw[fill=none, line width=0.05em] ({sin(50)*2*.5em},{.5em + cos(50)*2*.5em}) circle (0.2em and 0.5em);
				\draw[fill=none, line width=0.05em] ({-sin(50)*2*.5em},{.5em + cos(50)*2*.5em}) circle (0.2em and 0.5em);

				\draw[fill=none, line width=0.05em] ({0},{3.5*.5em + cos(50)*2*.5em}) circle (0.5em and 0.2em);
				\draw[fill=black] (-.5em,{3.5*.5em + cos(50)*2*.5em}) circle (0.10em and 0.10em);
				\draw[fill=black] (+.5em,{3.5*.5em + cos(50)*2*.5em}) circle (0.10em and 0.10em);
				\draw[fill=none, line width=0.05em, rotate around={+50:(-.5em,{3.5*.5em + cos(50)*2*.5em})}] (-.5em,{4.5*.5em + cos(50)*2*.5em}) circle (0.2em and 0.5em);
				\draw[fill=none, line width=0.05em, rotate around={-50:(+.5em,{3.5*.5em + cos(50)*2*.5em})}] (+.5em,{4.5*.5em + cos(50)*2*.5em})circle (0.2em and 0.5em);
				\node at ({.8*sin(50)*2*.5em},{1.5em + cos(50)*2*.5em}) {$.$};
				\node at ({.9*sin(50)*2*.5em},{1.2em + cos(50)*2*.5em}) {$.$};
				\node at ({-.8*sin(50)*2*.5em},{1.5em + cos(50)*2*.5em}) {$.$};
				\node at ({-.9*sin(50)*2*.5em},{1.2em + cos(50)*2*.5em}) {$.$};

				\draw[fill=none, line width=0.05em] (+.65em,-0.10em) -- (+0.85em,+0.10em);
				\draw[fill=none, line width=0.05em] (+.65em,+0.10em) -- (+0.85em,-0.10em);

				\draw[fill=none, line width=0.05em, rotate around={-45:(.65em,.30em)}] (0.65em-0.10em,.30em-0.10em) -- (0.65em+.10em,.30em+0.10em);
				\draw[fill=none, line width=0.05em, rotate around={-45:(.65em,.30em)}] (0.65em-0.10em,.30em+0.10em) -- (0.65em+.10em,.30em-0.10em);

				\draw[fill=none, line width=0.05em, rotate around={-45:(-.65em,.30em)}] (-0.65em-0.10em,.30em-0.10em) -- (-0.65em+.10em,.30em+0.10em);
				\draw[fill=none, line width=0.05em, rotate around={-45:(-.65em,.30em)}] (-0.65em-0.10em,.30em+0.10em) -- (-0.65em+.10em,.30em-0.10em);

				\draw[fill=none, line width=0.05em, rotate around={-45:(.45em,.60em)}] (0.45em-0.10em,.60em-0.10em) -- (0.45em+.10em,.60em+0.10em);
				\draw[fill=none, line width=0.05em, rotate around={-45:(.45em,.60em)}] (0.45em-0.10em,.60em+0.10em) -- (0.45em+.10em,.60em-0.10em);

				\draw[fill=none, line width=0.05em, rotate around={-45:(-.45em,.60em)}] (-0.45em-0.10em,.60em-0.10em) -- (-0.45em+.10em,.60em+0.10em);
				\draw[fill=none, line width=0.05em, rotate around={-45:(-.45em,.60em)}] (-0.45em-0.10em,.60em+0.10em) -- (-0.45em+.10em,.60em-0.10em);

				\draw[fill=none, line width=0.05em] (-1.25em-0.10em,.85em-0.10em) -- (-1.25em+.10em,.85em+0.10em);
				\draw[fill=none, line width=0.05em] (-1.25em-0.10em,.85em+0.10em) -- (-1.25em+.10em,.85em-0.10em);

				\draw[fill=none, line width=0.05em] (1.25em-0.10em,.85em-0.10em) -- (1.25em+.10em,.85em+0.10em);
				\draw[fill=none, line width=0.05em] (1.25em-0.10em,.85em+0.10em) -- (1.25em+.10em,.85em-0.10em);

				\draw[fill=none, line width=0.05em] (0.97em-0.10em,1.15em-0.10em) -- (0.97em+.10em,1.15em+0.10em);
				\draw[fill=none, line width=0.05em] (0.97em-0.10em,1.15em+0.10em) -- (0.97em+.10em,1.15em-0.10em);

				\draw[fill=none, line width=0.05em] (-0.97em-0.10em,1.15em-0.10em) -- (-0.97em+.10em,1.15em+0.10em);
				\draw[fill=none, line width=0.05em] (-0.97em-0.10em,1.15em+0.10em) -- (-0.97em+.10em,1.15em-0.10em);

				\draw[fill=none, line width=0.05em] (0.57em-0.10em,1.15em-0.10em) -- (0.57em+.10em,1.15em+0.10em);
				\draw[fill=none, line width=0.05em] (0.57em-0.10em,1.15em+0.10em) -- (0.57em+.10em,1.15em-0.10em);

				\draw[fill=none, line width=0.05em] (-0.57em-0.10em,1.15em-0.10em) -- (-0.57em+.10em,1.15em+0.10em);
				\draw[fill=none, line width=0.05em] (-0.57em-0.10em,1.15em+0.10em) -- (-0.57em+.10em,1.15em-0.10em);

				\draw[fill=none, line width=0.05em, rotate around={-45:(-0.97em,2.52em)}] (-0.97em-0.10em,2.52em-0.10em) -- (-0.97em+.10em,2.52em+0.10em);
				\draw[fill=none, line width=0.05em, rotate around={-45:(-0.97em,2.52em)}] (-0.97em-0.10em,2.52em+0.10em) -- (-0.97em+.10em,2.52em-0.10em);

				\draw[fill=none, line width=0.05em, rotate around={-45:(0.97em,2.52em)}] (0.97em-0.10em,2.52em-0.10em) -- (0.97em+.10em,2.52em+0.10em);
				\draw[fill=none, line width=0.05em, rotate around={-45:(0.97em,2.52em)}] (0.97em-0.10em,2.52em+0.10em) -- (0.97em+.10em,2.52em-0.10em);

				\draw[fill=none, line width=0.05em] (0.em-0.10em,2.20em-0.10em) -- (0.em+.10em,2.20em+0.10em);
				\draw[fill=none, line width=0.05em] (0.em-0.10em,2.20em+0.10em) -- (0.em+.10em,2.20em-0.10em);
				\draw[fill=none, line width=0.05em] (0.em-0.10em,2.60em-0.10em) -- (0.em+.10em,2.60em+0.10em);
				\draw[fill=none, line width=0.05em] (0.em-0.10em,2.60em+0.10em) -- (0.em+.10em,2.60em-0.10em);
			\end{tikzpicture}
			+  \begin{tikzpicture}[baseline=1.7em,scale=1.3]
				
				\draw[fill=none, line width=0.05em, rotate around={-50:(1.05em,.05em)}] (1.05em-0.10em,.05em-0.10em) -- (01.05em+.10em,.05em+0.10em);
				\draw[fill=none, line width=0.05em, rotate around={-50:(1.05em,.05em)}] (1.05em-0.10em,.05em+0.10em) -- (01.05em+.10em,.05em-0.10em);

				\draw[fill=black] ({sin(50)*2*.5em},{cos(50)*2*.5em}) circle (0.10em and 0.10em);
				\draw[fill=black] ({-sin(50)*2*.5em},{cos(50)*2*.5em}) circle (0.10em and 0.10em);
				\draw[fill=none, line width=0.05em]  (-1.2em,-0.2em) -- ({-sin(50)*2*.5em},{cos(50)*2*.5em}) -- ({sin(50)*2*.5em},{cos(50)*2*.5em}) -- (+1.2em,-0.2em);

				\draw[fill=none, line width=0.05em] ({.5em + sin(50)*2*.5em},{cos(50)*2*.5em}) circle (0.5em and 0.2em);
				\draw[fill=none, line width=0.05em] ({-.5em - sin(50)*2*.5em},{cos(50)*2*.5em}) circle (0.5em and 0.2em);

				\draw[fill=none, line width=0.05em] ({sin(50)*2*.5em},{.5em + cos(50)*2*.5em}) circle (0.2em and 0.5em);
				\draw[fill=none, line width=0.05em] ({-sin(50)*2*.5em},{.5em + cos(50)*2*.5em}) circle (0.2em and 0.5em);

				\draw[fill=none, line width=0.05em] ({0},{3.5*.5em + cos(50)*2*.5em}) circle (0.5em and 0.2em);
				\draw[fill=black] (-.5em,{3.5*.5em + cos(50)*2*.5em}) circle (0.10em and 0.10em);
				\draw[fill=black] (+.5em,{3.5*.5em + cos(50)*2*.5em}) circle (0.10em and 0.10em);
				\draw[fill=none, line width=0.05em, rotate around={+50:(-.5em,{3.5*.5em + cos(50)*2*.5em})}] (-.5em,{4.5*.5em + cos(50)*2*.5em}) circle (0.2em and 0.5em);
				\draw[fill=none, line width=0.05em, rotate around={-50:(+.5em,{3.5*.5em + cos(50)*2*.5em})}] (+.5em,{4.5*.5em + cos(50)*2*.5em})circle (0.2em and 0.5em);
				\node at ({.8*sin(50)*2*.5em},{1.5em + cos(50)*2*.5em}) {$.$};
				\node at ({.9*sin(50)*2*.5em},{1.2em + cos(50)*2*.5em}) {$.$};
				\node at ({-.8*sin(50)*2*.5em},{1.5em + cos(50)*2*.5em}) {$.$};
				\node at ({-.9*sin(50)*2*.5em},{1.2em + cos(50)*2*.5em}) {$.$};

				\draw[fill=none, line width=0.05em] (-1.25em-0.10em,.85em-0.10em) -- (-1.25em+.10em,.85em+0.10em);
				\draw[fill=none, line width=0.05em] (-1.25em-0.10em,.85em+0.10em) -- (-1.25em+.10em,.85em-0.10em);

				\draw[fill=none, line width=0.05em] (1.25em-0.10em,.85em-0.10em) -- (1.25em+.10em,.85em+0.10em);
				\draw[fill=none, line width=0.05em] (1.25em-0.10em,.85em+0.10em) -- (1.25em+.10em,.85em-0.10em);

				\draw[fill=none, line width=0.05em] (0.97em-0.10em,1.15em-0.10em) -- (0.97em+.10em,1.15em+0.10em);
				\draw[fill=none, line width=0.05em] (0.97em-0.10em,1.15em+0.10em) -- (0.97em+.10em,1.15em-0.10em);

				\draw[fill=none, line width=0.05em] (-0.97em-0.10em,1.15em-0.10em) -- (-0.97em+.10em,1.15em+0.10em);
				\draw[fill=none, line width=0.05em] (-0.97em-0.10em,1.15em+0.10em) -- (-0.97em+.10em,1.15em-0.10em);

				\draw[fill=none, line width=0.05em] (0.57em-0.10em,1.15em-0.10em) -- (0.57em+.10em,1.15em+0.10em);
				\draw[fill=none, line width=0.05em] (0.57em-0.10em,1.15em+0.10em) -- (0.57em+.10em,1.15em-0.10em);

				\draw[fill=none, line width=0.05em] (-0.57em-0.10em,1.15em-0.10em) -- (-0.57em+.10em,1.15em+0.10em);
				\draw[fill=none, line width=0.05em] (-0.57em-0.10em,1.15em+0.10em) -- (-0.57em+.10em,1.15em-0.10em);

				\draw[fill=none, line width=0.05em, rotate around={-45:(-0.97em,2.52em)}] (-0.97em-0.10em,2.52em-0.10em) -- (-0.97em+.10em,2.52em+0.10em);
				\draw[fill=none, line width=0.05em, rotate around={-45:(-0.97em,2.52em)}] (-0.97em-0.10em,2.52em+0.10em) -- (-0.97em+.10em,2.52em-0.10em);

				\draw[fill=none, line width=0.05em, rotate around={-45:(0.97em,2.52em)}] (0.97em-0.10em,2.52em-0.10em) -- (0.97em+.10em,2.52em+0.10em);
				\draw[fill=none, line width=0.05em, rotate around={-45:(0.97em,2.52em)}] (0.97em-0.10em,2.52em+0.10em) -- (0.97em+.10em,2.52em-0.10em);

				\draw[fill=none, line width=0.05em] (0.em-0.10em,2.20em-0.10em) -- (0.em+.10em,2.20em+0.10em);
				\draw[fill=none, line width=0.05em] (0.em-0.10em,2.20em+0.10em) -- (0.em+.10em,2.20em-0.10em);
				\draw[fill=none, line width=0.05em] (0.em-0.10em,2.60em-0.10em) -- (0.em+.10em,2.60em+0.10em);
				\draw[fill=none, line width=0.05em] (0.em-0.10em,2.60em+0.10em) -- (0.em+.10em,2.60em-0.10em);

				\draw[fill=none, line width=0.05em] (0.em-0.10em,{cos(50)*2*.5em-0.10em}) -- (0.em+.10em,{cos(50)*2*.5em+0.10em});
				\draw[fill=none, line width=0.05em] (0.em-0.10em,{cos(50)*2*.5em+0.10em}) -- (0.em+.10em,{cos(50)*2*.5em-0.10em});

			\end{tikzpicture},
	\end{equation}
	where the small dots denote arbitrarily many insertions of the same shape, and~\tikz [baseline=-0.3em] {
		\draw[fill=white, line width=0.05em] (0em,0em) circle (0.3em and 0.3em);
		\draw[fill=none,  line width=0.05em] (0em+.15em,0em+.15em) -- (0em-.15em,0em-.15em);
		\draw[fill=none,  line width=0.05em] (0em+.15em,0em-.15em) -- (0em-.15em,0em+.15em);
	}~denotes the sextic tree-level counterterm insertion. The last family of diagrams in~\eq{gap-BMB-NLO} yields an explicit dependence on the external momentum. Hence, the gap equation at NLO does not merely describe a dynamical mass parameter but rather a more complicated dressed propagator. 
	
	Setting aside this complication, the dynamical mass parameter $\mathscr{M}$ gains additional contributions with respect to the LO condition~\eq{gap-BMB} 
	\begin{equation}
		\mathscr{M}^2 = \mathscr{M}^2 \left[\bar{\eta} + \beta_{\bar{\eta}}\log \frac{\mathscr{M}}{\mu} + \dots\right]\,,
	\end{equation}
	which only permits the solution $\mathscr{M} \propto \mu$ as well as the ground state $\mathscr{M} = 0$.  

	As $V_\mathscr{M}$ is not a flat function of $\mathscr{M}$, the composite field \eq{Dilaton} ceases to be a dilaton. We will not compute explicit corrections to its LO mass~\eq{Dilaton-mass} as it will vanish due to $\mathscr{M} = 0$. Note that the authors of~\cite{Omid:2016jve} have employed saddle point methods to argue that for $\mathscr{M} \neq 0$ the mass is tachyonic at $\bar{\eta} = 1$. A cross-check of this result is beyond the scope of this work.

\section{Conclusions}\label{sec:Conclusion}

In this work, we revisited the $O(N)$ model with $\phi^6$ interactions using perturbation theory, dimensional regularisation, and minimal subtraction. Through resummations, we derived exact expressions in a systematic $1/N$ expansion, which we used as a guardrail into strongly coupled regimes.
We also employed composite-operator effective-action techniques to track possible competing vacua.
This set of tools allowed us to reconcile and extend various literature results, including those obtained via saddle-point approximations or the FRG.

Each of our findings supports the conclusion that the UV fixed point found in~\cite{Townsend:1976sy,Appelquist:1981sf,Pisarski:1982vz} exists and is accessible within a weakly coupled regime for realistic QFTs at large $N$.
Concretely, we demonstrated that the fixed point becomes vanishingly small in a $N\to \infty$ limit, which ensures its persistence to higher-loop orders in the $\beta$-function. This is an extension to the arguments brought forth in~\cite{Appelquist:1981sf,Pisarski:1982vz}, which leaves no room for doubt about the viability of the perturbative expansion.
Moreover, we computed the first complete expression for the effective potential at NLO in the $1/N$-expansion, superseding any partial attempts in prior literature~\cite{Townsend:1975kh,Matsubara:1984rk}. 
Our results suggest that for sufficiently large $N$, the fixed point is stable as spontaneous breaking of the global $O(N)$ symmetry does not occur.
Furthermore, we showed that the tricritical line of conformality does not interfere with the existence of the UV fixed point; its well-known disappearance~\cite{David:1984we,David:1985zz,Amit:1984ri,Yabunaka:2017uox,Fleming:2020qqx} is the consequence of a consistent $1/N$-expansion.
Finally, we brought forward new arguments which show the absence of the BMB instability~\cite{Bardeen:1983rv} at NLO. Thus, there is no evidence for any obstructions to the UV fixed point.

Our findings suggest that both the tricritical line of conformality, as well as the BMB phenomenon at its endpoint, are ephemeral products of a strict $N\to\infty$ limit. They are based on an accidental scale invariance only present at LO in the $1/N$-expansion, which is broken by quantum fluctuations in realistic large-$N$ QFTs. 
We interpret this LO scale invariance as a relict of the classical scale symmetry.
Once quantum corrections are fully accounted for, the scale invariance is broken and only restored at the UV fixed point~\cite{Townsend:1976sy,Appelquist:1981sf,Pisarski:1982vz}. 

We expect that the results of this paper, including the persistence of Pisarski's fixed point, can be corroborated using the FRG, provided a sufficiently nuanced truncation is employed. We leave this exploration to future work.

\begin{center}\small\textbf{\uppercase{Acknowledgements}}\end{center}

We are indebted to E. Stamou for comments on the manuscript.
We thank C.~Cresswell--Hogg, D.~F.~Litim,  M.~M.~Scherer, and Y.~Schröder for discussions.
M.U. is supported by the doctoral scholarship program of the \textit{Studienstiftung des deutschen Volkes} and the Mercator Research Center Ruhr under Project No.~Ko-2022-0012. 
The work of S.K. has been funded by Consejería de Universidad, Investigación e Innovación, Gobierno de España and Unión Europea – NextGenerationEU under grants AST22 6.5 and CNS2022-136024 and by MICIU/AEI/10.13039/501100011033 and FEDER/UE (grant PID2022-139466NB-C21).

%\vspace*{10em}

\begin{widetext}

\appendix

\section{Details on Multiloop Integrals in Large \texorpdfstring{$N$}{N}}\label{sec:appendix}

In this Appendix, we collect the relevant techniques and provide further details on the multiloop integrals that appear in the main text.
We always work in the minimal subtraction scheme and regularise loop integrals in $d=3-2\varepsilon$ dimensions. 
To evaluate the integrals, we employ the \texttt{MaRTIn}~\cite{Brod:2024zaz} framework, which in turn uses 
\texttt{QGRAF}~\cite{Nogueira:1991ex} for diagram generation and \texttt{FORM}~\cite{Kuipers:2012rf} for the symbolic computations.

As we compute the effective potential at leading order (LO) and next-to-leading order (NLO) in a $1/N$ expansion, an infinite amount of loop diagrams
has to be resummed.
The LO contributions take the form of tadpoles~\eq{V1PI-LO} which factorise and are finite. They can be resummed by introducing a dynamical mass parameter that can be determined via a gap equation~\eq{gap-LO}.
Explicitly, the tadpoles $\mathcal{T}$ are finite and read
\begin{equation}\label{eq:tadpoles-app}
	\mathcal{T} =
	\begin{tikzpicture}[baseline=-.3em]
		\draw[fill=none, line width=0.05em] (0em,+.6em) circle (0.6em and 0.6em);
		\draw[fill=black] (0,0) circle (0.15em and 0.15em);
	\end{tikzpicture}
	= \int \frac{\mathrm{d}^d k}{(2\pi)^d}\frac{1}{k^2-m^2}\,= - \frac{i |m|}{(4\pi)^{3/2}} \left(\frac{\mu^2}{|m|^2} e^{\gamma_E} \right)^{\varepsilon} \Gamma\left(\varepsilon-\frac12 \right)  = \frac{i |m|}{4 \pi} + \mathcal{O}(\varepsilon) \,.
\end{equation}

At NLO, there are sets of ring diagrams $R_n$~\eq{Rn} and $\tilde{R}_{n}$~\eq{Rtilden-dia}, which cannot be resummed by the same means. They incorporate chains of
subgraphs~\eq{dressed-bubble}, which contain one-loop bubble diagrams that explicitly depend on the momentum routed through them from external legs.
In general $d$ dimensions, each of these bubbles $\mathcal{B}$ reads
\begin{equation}
	\mathcal{B} = \tikz [baseline=-0.3em] {

	\draw[fill=black] (+0.8em,0em) circle (0.15em and 0.15em);
	\draw[fill=black] (-0.8em,0em) circle (0.15em and 0.15em);
	\draw[fill=none, line width=0.05em] (0em,0em) circle (0.8em and 0.8em);	
} 
= \int \frac{\mathrm{d}^d k}{(2\pi)^d}\frac{1}{(k^2-m^2)((k+p)^2-m^2)} \,.
\end{equation}
Using Feynman parametrisation, Wick rotation and standard integration methods, we arrive at
\begin{equation}\label{eq:bubble-B}
	\mathcal{B}=\frac{i}{4 \pi p_{E}} \arctan\left( \frac{p_E}{2m} \right) +  \mathcal{O}(\varepsilon)\,,
\end{equation}
where $p_E$ is the Euclidean momentum. We find this expression to be in agreement with \cite{Kotikov:2021hsy}. 

The diagrams $\tilde{R}_{1,1}$~\eq{R1}, $R_2$~\eq{R2} and $R_3$~\eq{R3} are UV divergent and can be directly obtained by combining the respective two-loop~\cite{Davydychev:1992mt}, three-loop~\cite{Rajantie:1996np} and four-loop integrals~\cite{Lee:2010hs} and factorising LO tadpoles~\eq{tadpoles-app}.
The remaining diagrams $R_{n\geq4}$~\eq{Rn_sol}, $\tilde{R}_{1,l\geq2}$~\eq{Rtilde1n} and $\tilde{R}_{n\geq2}$~\eq{Rtilden} are finite. We have opted to sum them as the common expression $\sum_{n=4}^\infty R_n + \sum_{n=2}^\infty \left(\tilde{R}_{1,l} + \tilde{R}_{n}\right)$ in~\eq{FF-sum}, introducing the function $F(x)$ as defined in~\eq{FF}, which encapsulates the integration over the bubble momentum $p_E$ from \eq{bubble-B}. In the effective potential at NLO~\eq{Veff-NLO}, the evaluation of $F(x)$ is performed numerically.

To obtain the result for $V_\mathscr{M}$~\eq{VM-M}, merely a resummation of $R_{n\geq4}$ is required, which is abbreviated as \eq{RnLarger4} introducing the function $f(x)$ in the same manner as before~\eq{ff1}. This function $f(x)$ has a finite the radius of convergence for positive arguments $x\leq1$, which stems from the sum $\sum_n R_n$ being divergent. Thus, $f(x)$ is only well-defined within this radius of convergence, as it relies on exchanging the summation of diagrams and individual loop integrations.
As before, $f(x)$ is evaluated numerically to discuss~\eq{VM-M}.

While we have found the numerical evaluation of the summed expression $\sum_{n=4}^\infty R_n$~\eq{RnLarger4} to be more practical, we could have chosen to compute each of the contributions $R_n$~\eq{Rn_sol} analytically. For $n\geq4$ they can be written as 
\begin{equation}\label{eq:Rn_In}
	R_n = - \frac{2 M^3}{\pi^2 n } \left[\bar{\eta} \left(1- \frac{4\pi \varphi^2}{N M} \right)\right]^n I_n \,,
\end{equation}
where the integral $I_n$ defined as
\begin{equation}
	I_n =\int_0^\infty \!\!\!  \frac{\arctan^n(z)}{z^{n-2}}\mathrm{d} z
\end{equation}
can be obtained recursively. After repeated integration by parts, we can write down $I_n$ in terms of base integrals $J_n$, which are defined below and can be integrated more easily. As expressions differ slightly depending on whether $n$ is odd or even, we use superscripts $I_n^{\text{odd}}$ and $I_n^{\text{even}}$ to discriminate between the two cases. The integral $I_n$ can reduced in a recursion 
\begin{align}
	I_n^{\text{odd}}&=\frac{n}{n-3}\left[I_{n-1}^{\text{even}}+(-1)^{(n-1)/2}J_{n-1}^{\text{even}}+\mathcal{S}_{n-1}^{\text{even}}(n-5)\right]\,, \qquad n>5\,,\\
	I_n^{\text{even}}&=\frac{n}{n-3}\left[I_{n-1}^{\text{odd}}+(-1)^{n/2}J_{n-1}^{\text{odd}}+\mathcal{S}_{n-1}^{\text{odd}}(n-5)\right]\,, \qquad \qquad n>4\,,
\end{align}
which is terminated at the lowest cases of $n$, where the integrals read
\begin{align}
	I_4=4 J_3\,, \qquad\text{ and }\qquad  I_5=\frac{5}{2}J_4\,.
\end{align}
Here, $\mathcal{S}_n$ are recursive sums defined via
\begin{align}
	\mathcal{S}_n^{\text{odd}}(m)&=\sum^{(m-1)/2}_{k=1}\frac{(-1)^{(m-2k-1)/2}n}{2k}\left((-1)^k J_{n-1}^{\text{even}}+\mathcal{S}_{n-1}^{\text{even}}(2k)\right)\,, \qquad m\geq 3\,, \\
	\mathcal{S}_n^{\text{even}}(m)&=\sum^{m/2}_{k=2}\frac{(-1)^{(m-2k)/2}n}{2k-1}\left((-1)^k J_{n-1}^{\text{odd}}+\mathcal{S}_{n-1}^{\text{odd}}(2k-1)\right)\,,\qquad\,\quad m\geq 4\,.
\end{align}
After performing all recursions, only base integrals $J_n$ remain, which can be computed directly and are defined as
\begin{equation}
	J_n^\text{odd}=\int_0^\infty \!\!\!  \frac{\arctan^n(z)}{z(1+z^2)}\mathrm{d} z=\left(\frac{\pi}{2}\right)^n\log 2+\sum_{m=1}^{(n-1)/2}(-1)^m\left(\frac{\pi}{2}\right)^{n-2m}\frac{(2^{2m}-1)}{2^{4m}}\frac{n!\zeta_{2m+1}}{(n-2m)!}\,,
\end{equation}
for odd values of $n$, whereas $J_n$ for even-$n$ is related to the $J_n^\text{odd}$  by
\begin{equation}
		J_n^\text{even}=\int_0^\infty \!\!\!  \frac{\arctan^n(z)}{z^2(1+z^2)}\mathrm{d} z= -\frac{1}{(n+1)}\left(\frac{\pi}{2}\right)^{n+1}+n J_{n-1}^\text{odd}\,.
\end{equation}

\end{widetext}

\FloatBarrier
\bibliography{ref.bib}
\bibliographystyle{JHEP}
 
\end{document}